\documentclass[a4paper,fleqn]{cas-sc}

\usepackage[numbers]{natbib}

\def\tsc#1{\csdef{#1}{\textsc{\lowercase{#1}}\xspace}}
\tsc{WGM}
\tsc{QE}
\tsc{EP}
\tsc{PMS}
\tsc{BEC}
\tsc{DE}

\usepackage{amsmath}
\usepackage{lineno}
\usepackage{url}
\usepackage{float}
\usepackage{subcaption}
\usepackage{algorithm}
\usepackage{algpseudocode}
\usepackage{longtable}
\usepackage{nomencl}

\makenomenclature

\usepackage{lineno}

\usepackage{xcolor, soul}
\sethlcolor{yellow}

\newcommand{\tenone}[1]{{\ensuremath{\boldsymbol{#1}}}}
\newcommand*{\tentwo}[1]{\tenone{\tenone{#1}}}

\newcommand*{\tenfour}[1]{\tentwo{\tentwo{#1}}}
\renewcommand*{\div}[1]{\nabla\cdot({#1})}
\newcommand*{\x}{\tenone{x}}

\newcommand*{\stress}{\tentwo{\sigma}}
\newcommand*{\strain}{\tentwo{\epsilon}}
\newcommand*{\CC}{\tenfour{C}}
\newcommand*{\Stress}{\langle \tentwo{\sigma} \rangle}
\newcommand*{\Strain}{\langle \tentwo{\epsilon} \rangle }

\begin{document}
\let\WriteBookmarks\relax
\def\floatpagepagefraction{1}
\def\textpagefraction{.001}
\shorttitle{Numerical predictions of cemented granular medium during debonding}
\shortauthors{A. Sac-Morane et~al.}

\title[mode = title]{Predicting the Elastic Properties of a Cemented Granular Material during Chemical Damage (Debonding)}                      

\author[1,2,3]{A. Sac-Morane}[orcid=0009-0008-5454-8107]
\cormark[1]
\ead{alexandre.sac-morane@enpc.fr}
\ead[url]{https://alexsacmorane.github.io/}
\credit{Conceptualization, Formal analysis, Investigation, Methodology, Software, Validation, Visualization, Writing – original draft}

\affiliation[1]{organization={Navier, CNRS, Université Gustave Eiffel, ENPC, Institut Polytechnique de Paris},
                city={Marne-la-Vallée},
                country={France}}
\affiliation[2]{organization={Multiphysics Geomechanics Lab, Duke University},
                addressline={Hudson Hall Annex, Room No. 053A}, 
                postcode={27708}, 
                city={Durham},
                state={NC},
                country={USA}}
\affiliation[3]{organization={Institute of Mechanics, Materials and Civil Engineering, UCLouvain},
                addressline={Place du Levant 1}, 
                postcode={1348}, 
                city={Louvain-la-Neuve},
                country={Belgium}}
\cortext[cor1]{Corresponding author}

\author[1]{M. Olarte-Garzon}[orcid=0000-0003-0253-5618]
\credit{Formal analysis, Investigation, Methodology, Software, Validation, Writing – original draft}
\ead{maria-camila.olarte-garzon@enpc.fr}

\author[1]{P. Braun}[orcid=0000-0003-3207-5018]
\ead{philipp.braun@enpc.fr}
\ead[url]{https://navier-lab.fr/equipe/braun-philipp/}
\credit{Funding acquisition, Supervision, Writing – review \& editing}

\author[1]{J.-M. Pereira}[orcid=0000-0002-0290-5191]
\ead{jean-michel.pereira@enpc.fr}
\ead[url]{https://navier-lab.fr/equipe/pereira-jean-michel/}
\credit{Funding acquisition, Supervision, Writing – review \& editing}

\author[2]{M. Veveakis}[orcid=0000-0002-4911-6026]
\ead{manolis.veveakis@duke.edu}
\ead[url]{https://cee.duke.edu/people/manolis-veveakis/}
\credit{Conceptualization, Funding acquisition, Supervision, Writing – review \& editing}

\author[3]{H. Rattez}[orcid=0000-0002-7245-6563]
\ead{hadrien.rattez@uclouvain.be}
\ead[url]{https://www.uclouvain.be/en/people/hadrien.rattez}
\credit{Conceptualization, Funding acquisition, Supervision, Writing – review \& editing}

\begin{abstract}
While underground reservoirs emerge as essential elements to face global warming, these systems represent complex multi-physical and multiscale problems.
The considered injection of fluids during hydrogen storage, carbon dioxide sequestration, or geothermal energy recovery involves a modification of the chemical equilibrium of the fluid in the porous reservoir.
Chemical reactions can induce microstructural changes of the rock matrix, leading to a reduction of elastic properties of the material, and to potential settlement or stress redistribution.
Consequently, it becomes pivotal to establish predictive behavior laws to describe the effect of chemical damage on elastic properties.

Facing the difficulties to estimate experimentally the impact of chemical damage on mechanical properties, a Digital Rock Physics approach is proposed in this contribution. 
This numerical homogenization scheme is used to compare two distinct types of microstructure models: the first one consists in a Discrete Element Model, while the second one employs a continuous description.
This continuous formulation is based on a Phase-Field description to predict the evolution of the microstructure subjected to chemical alterations and on the Fast Fourier Transform to estimate the macroscopic properties of the material.
Finally, these frameworks establish different softening laws that can be used as constitutive ingredients for a cemented material during its weathering.
\end{abstract}


\maketitle

\section{Introduction}
\label{sec intro}

Renewable energy sources, such as wind and solar power, are required to reduce the impact of human activity on climate evolution.
However, due to their high intermittency, strategies for large-scale energy storage must be developed. 
For instance, the excess energy can be transformed into hydrogen (H$_\mathrm{2}$), which is then stored until its extraction and consumption during periods of energy deficiency.
In this context, underground storage in porous formations and salt caverns appears as a promising solution \citep{Heinemann2021,Lesueur2023}, offering a greater storage capacity than conventional aboveground infrastructure.
Likewise, geothermal energy can be employed as a renewable energy production method, injecting a fluid underground to extract the thermal energy of the Earth \citep{McCartney2016,Barbier2002}.
Moreover, deep aquifers have emerged as promising candidates for long-term sequestration of greenhouse gases, including carbon dioxide (CO$_\mathrm{2}$) \citep{Kanin2024,Liu2026}. If the geothermal energy production is already a mature engineering application and the feasibility of the CO$_\mathrm{2}$ sequestration has been proven through commercial sites, the viability of the H$_\mathrm{2}$ storage is currently explored through various pilot projects. 

These engineering applications involve the injection/extraction of a fluid into a porous reservoir. Such an operation modifies the fluid chemical composition, and thus its reactivity with the surrounding solid phases, activating or enhancing chemical reactions \citep{Liteanu2012,Lesueur2020}.
This chemical destabilization involves multiple consequences due to the multiphysical aspect of the reservoir behavior: settlement at the surface \citep{LeGuen2007,Brzesowsky2014,SacMorane:PFDEM,SacMorane:PFDEMb}, modifications of the mechanical \citep{Lesueur2023,Wang2016,Nova2003,SacMorane:StressState, Qiao2026} and hydraulic \citep{Hueckel2005,Vallin2013,Lesueur2020b,Soulaine2021} properties, or threat to the stability of the caprock/reservoir \citep{Rohmer2016, Manceau2016,Stefanou2014}. It has even been correlated to earthquake nucleation \citep{Blocher2018,Rattez2020,Rattez2021,Lengline2023}.
Similar processes are observed at the surface during natural weathering of rock \citep{Palmer1973,Vaughan1984}.
Even if the coupling between the mechanics and the chemistry is not the sole relation involved in the reservoir problem, the scope of this paper is limited to it.

\begin{figure}[h]
    \centering
    \includegraphics[width=0.8\linewidth]{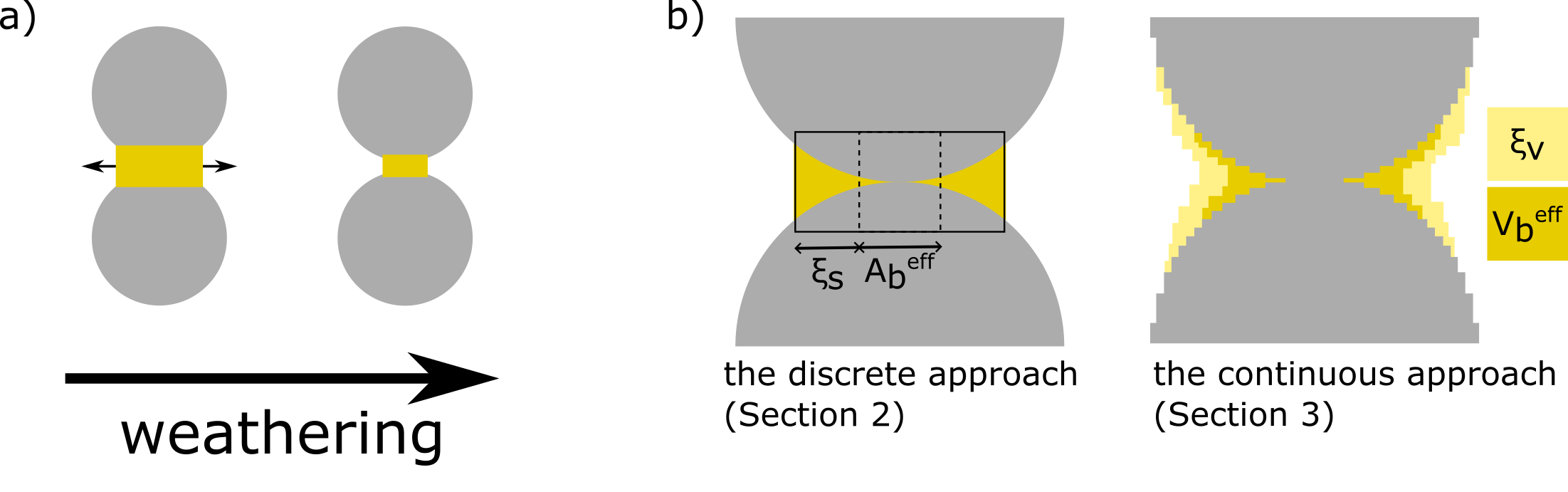}
    \caption{a) Schematic representation of the debonding phenomenon (the bonds dissolve while the grains remain intact) and b) models employed in this contribution (a cement surface reduction or a mass removal).}
    \label{weathering rock}
\end{figure}

In the context of reservoir rocks which are predominantly composed of bonded granular material, chemical damage can be conceptualized as three distinct phenomena.
The first chemical alteration is the transformation of a mineral into one with a weaker stiffness or lower density.
However, this process appears to occur mostly, for the shallow depth considered in this paper, during hydrothermal alteration in volcanic conditions \citep{Detienne2016}, which is not considered here.
The second phenomenon is the dissolution of grains and the bonds located between them \citep{Castellanza2004, Ciantia2014}. 
This degradation is exemplified by calcarenite, wherein the grains and bonds are composed of calcium carbonate, which both dissolve under chemical attack.
This type of weathering has been extensively investigated experimentally and numerically in the literature \citep{Castellanza2004, Ciantia2014, Shin2009, Buscarnera2012, Cha2014, Cha2016}.
The third case is the dissolution of the bonds only, while the grains remain intact \citep{Castellanza2004}. Depicted in Figure \ref{weathering rock}a, an illustrative example is provided by silicic sand, which contains carbonate bonds. The dissolution kinetics of the bond is assumed to be considerably faster than that of the grain. This phenomenon is referred to as debonding and is the subject of the present work. 

The impact of debonding, and weathering in general, on the mechanical behavior of geomaterials has been previously investigated through both experiments \citep{Ciantia2014, Ciantia2014b, Basu2009, Heap2021, Geindreau2022} and numerical simulations \citep{Nova2003, Stefanou2014, Gajo2015, Loret2002, Hu2013, Hu2019, Wu2025, DeSimone2025, Yang2026} in a range of configurations. However, even if the elastic properties of the reservoir material are fundamental to predict settlement behavior, only a limited number of studies have focused on examining them while this chemical degradation occurs. 
For instance, the impact of the chemistry on the mechanical behavior of rocks is restricted to the plasticity criterion and the elasticity parameters remain constant in \citep{Nova2003, Stefanou2014, Ciantia2013}.
Subsequently, \emph{Gajo et al.} introduced the dependence of the mechanical performance of the material on the chemical damage in \citep{Gajo2015, Gajo2019}.
However, their model is based on a simplistic microstructure with a cubic arrangement of monodisperse grains, and several pivotal mechanisms occurring at the microscale, such as the granular reorganization, are not considered.
Similarly, \emph{Buscarnera and Das} developed a model based on a thermodynamic framework that is able to predict the evolution of the elastic parameter during debonding and breakage \citep{Buscarnera2016}. 
However, the considered microstructure remains simplified as the cement phase is not explicitly described, and the granular skeleton is depicted with a particle size distribution solely.
On the same note as earlier, this idealization does not capture all the mechanisms, such as the heterogeneity of the microstructure or the granular reorganization.

\begin{figure}[h]
    \centering
    \includegraphics[width=0.6\linewidth]{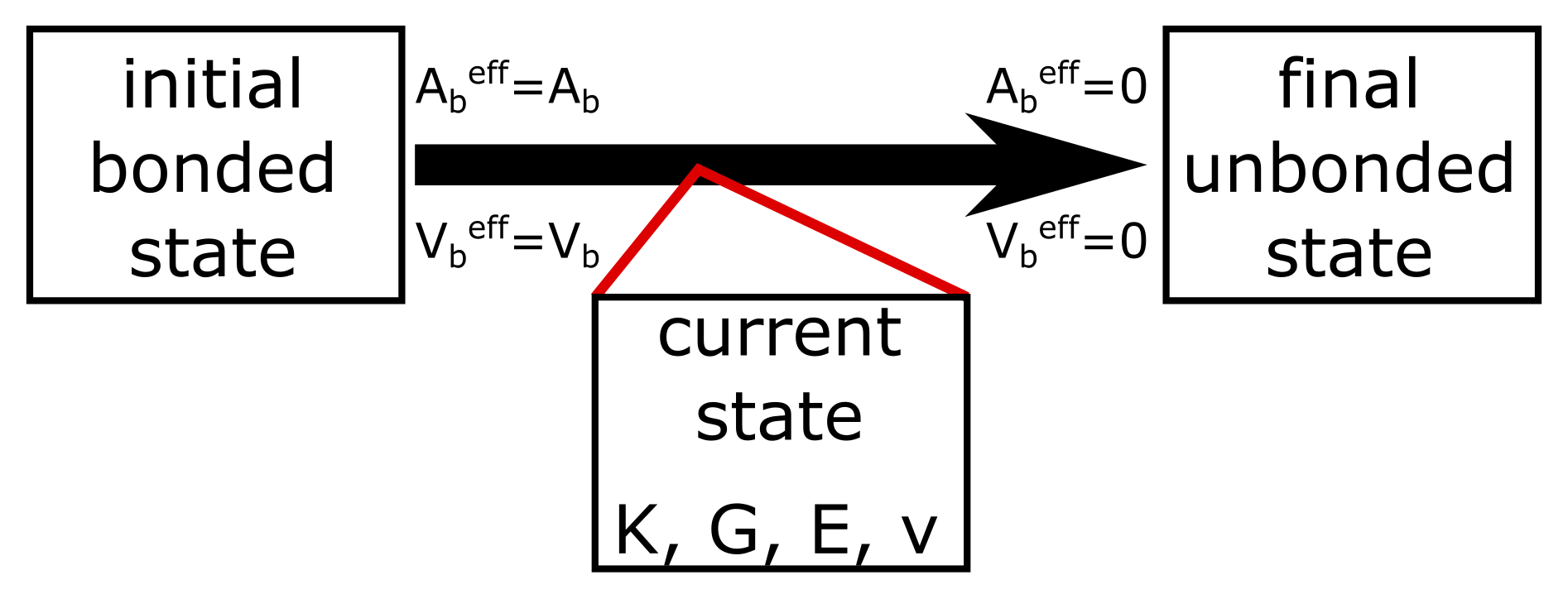}
    \caption{The chemical-induced evolution of the elastic parameters of the cemented granular medium is estimated by investigating multiple intermediate states between the intact rock and the unbonded material.}
    \label{Elastic Parameters vs Chemical Damage}
\end{figure}

The goal of this contribution is to fill the gap concerning the softening law of the elastic parameters of reservoir rocks subjected to chemical debonding.
To do so, a numerical framework is employed with the explicit description of the microstructure of a reservoir rock.
Subsequently, this formulation is used to estimate, in an explicit manner, its evolution during the debonding phenomenon.
Then, the variation of the elastic properties of the sample is determined with a numerical homogenization scheme considering intermediate states between the bonded and unbonded states, see Figure \ref{Elastic Parameters vs Chemical Damage}.
This operation transfers the influence of the mechanisms from the micro scale to the Representative Element Volume scale. 

In particular, this contribution proposes to evaluate quantitatively the performance of two formulations.
The first approach uses the Discrete Element Method (DEM) to replicate the mechanical behavior of the sample during its alteration.
Subsequently, the second approach consists in the use of a Phase-Field (PF) description to predict the evolution of the microstructure and in the application of the Fast Fourier Transform (FFT) to compute the properties of the altered rock.

In the same tone as previous numerical studies \citep{SacMorane:StressState,Sun2016} and experimental works \citep{Konstantinou2021, Wang2023}, the natural sandstones from underground reservoirs are represented as cemented granular material.
Section \ref{DEM Section} employs a calibrated model developed for biocemented sands \citep{Sarkis2022, Dadda2017}.
Similarly, Section \ref{PF+FFT Section} uses the scans of a sample made of cemented glass beads \citep{Tengattini2023, Tengattini2023b} to define the initial microstructure for the PF + FFT approach.
Unfortunately, the samples investigated in these two Sections differ due to a lack of data.

This paper is divided into four main sections. 
Firstly, Section \ref{DEM Section} details the use of the DEM for a cemented granular sample, while debonding occurs. 
Nevertheless, the DEM-based representation of the microstructure remains subject to improvement. Indeed, cementation at contacts is modeled through an additional contact stiffness rather than an explicit geometrical or mechanical description, which introduces a level of approximation.
Consequently, Section \ref{PF+FFT Section} depicts a framework where the PF method is employed to predict the microstructure (grain, cement, and pore) evolution during the debonding phenomenon.
The mechanical performance of the material are then determined with an FFT-based method applied to these predicted microstructures.
The softening laws obtained by the application of these two approaches in the context of debonding are depicted in Section \ref{Results Section}. 
Finally, a discussion is proposed in Section \ref{Discussion Section} on the use of these informed constitutive laws and on the comparison between the two formulations.


\section{The discrete element approach}
\label{DEM Section}

The Discrete Element Model is employed as a first approach to estimate the elastic properties of a cemented granular material during debonding, meaning dissolution of the bonds while the grains remain intact.
Indeed, this description has been successfully employed in numerous investigations on sand (unbonded material) \citep{Kawamoto2016,Wu2022,SacMorane:Rolling,Singh2024} and rocks (bonded material) \citep{Wang2008,Shen2016,Zhang2023,Sun2017}.

Here we consider the cement as an additional stiffness at the contact of the particles.
As depicted in Figure \ref{weathering rock}b, the amplitude of the additional rigidity is estimated from the bond material quantity characterized by the effective area of the bond $A_b^{eff}$.
Subsequently, the debonding phenomenon can be modeled by a reduction factor $\xi_s$ of this parameter.
Then, the impact of the debonding on the elastic properties of the material can be explored by the scheme depicted in Figure \ref{Elastic Parameters vs Chemical Damage}. 
This Section is based on the model described in \citep{Sarkis2022}, extended later in \citep{SacMorane:StressState}, which has been calibrated on experimental tests on biocemented sands.


\subsection{Theory and formulation}
\label{Theory DEM}

The DEM is an approach developed by \emph{Cundall and Strack} \citep{Cundall1979} for simulating granular materials at the particle level. 
The fundamental premise of this discrete approach is to explicitly consider the individual particles and their interactions within the material itself \citep{OSullivan2011}. Newton's laws (linear and angular momentum) are employed to compute the motion of the grains.
These balance equations incorporate the external (loading) and the internal (due to contacts) forces.
This seminal formulation was extended to rock materials by \citep{Potyondy2004}, adding a bond between the particles.
It appears that the cement is not explicitly described, but modeled as an additional stiffness and cohesion at the contacts.
Subsequently, the chemical degradation can be readily introduced by reducing the bond size which defines this rigidity and strength \citep{SacMorane:StressState}. 
These modifications destabilize the granular microstructure and affect the properties at the material scale. 

\vskip\baselineskip

The contact models between particles considered herein are governed by a cohesive law following \citep{Bourrier2013}. This cohesion permits the relative displacement between bonded particles. However, due to the presence of cementation, an additional stiffness is introduced. 
The normal, tangential, bending and twisting models are described in the following. 
The contact can be modeled as two springs in parallel, one representing the bond $K_{\cdot,b}$ and the other representing the unbonded grain-grain contact $K_{\cdot,u}$.


The internal force applied to the particle is defined as the sum of the normal force vectors $F_{ni}$ and the tangential force vectors $F_{si}$ over the contacts.
The normal force vector $F_{ni}$ is described as $F_{ni} = K_n\Delta_{ni}$, where $\Delta_{ni}$ is the normal overlap. An elastic stiffness $K_n=K_{n,b}+K_{n,u}$ is required, formulated in Equation \ref{Normal Stiffness Equation}.
Even if the two components can be estimated, only $K_n$ is used in practice.

\begin{equation}
    K_n = 2E_m\frac{R_1R_2}{R_1+R_2} \label{Normal Stiffness Equation}
\end{equation}
where $E_m$ is the contact Young modulus, $R_1$ and $R_2$ are the radii of the particles in contact. The Young modulus $E_m$ is defined in Equation \ref{Young Local}.

Similarly, the tangential force vector $F_{si}$ is described as $F_{si} = K_s \Delta_{si}$, where $\Delta_{si}$ is the tangential overlap. An elastic stiffness $K_s= K_{s,b}+K_{s,u}$ is required, formulated in Equation \ref{Tangential Stiffness Equation}.
\begin{equation}
    K_s = \nu_c K_n=2\nu_c E_m\frac{R_1R_2}{R_1+R_2} \label{Tangential Stiffness Equation}
\end{equation}
where $\nu_c$ is the contact Poisson ratio.

Correspondingly, the internal moment applied to the particle is defined as the sum of the bending moment vectors $M_{bi}$, the torque due to the tangential force vectors $F_{si}$, and the twisting moment vector $M_{ti}$ over the contacts including the particle.
The bending moment vector $M_{bi}$ is described as $M_{bi} = K_b\Delta\theta_{bi}$. An elastic stiffness $K_b=K_{b,b}+K_{b,u}$ is required, formulated in Equation \ref{Bending Stiffness Equation}.
\begin{equation}
    K_b = \alpha_b K_s R_1 R_2=2\alpha_b\nu_c E_m\frac{\left(R_1R_2\right)^2}{R_1+R_2} \label{Bending Stiffness Equation}
\end{equation}
where $\alpha_b$ is a non-dimensional factor, relating the bending and the tangential stiffnesses.

Similarly, the twisting moment vector $M_{ti}$ is described as $M_{ti} = K_t\Delta\theta_{ti}$. An elastic stiffness $K_t=K_{t,b}+K_{t,u}$ is required, formulated in Equation \ref{Twisting Stiffness Equation}.
\begin{equation}
    K_t = \alpha_t K_s R_1 R_2=2\alpha_t\nu_c E_m\frac{\left(R_1R_2\right)^2}{R_1+R_2} \label{Twisting Stiffness Equation}
\end{equation}
where $\alpha_t$ is a non-dimensional factor, relating the twisting and the tangential stiffnesses. The bending and the twisting resistances aim to reproduce the complex shape of the grain \citep{SacMorane:Rolling,Ai2011, Mollon2020} as spheres are used in this framework (numerically more efficient).

Furthermore, a Coulomb friction limit is introduced at the contact level, see Equation \ref{Bond criteria}. 
The bond will exist until one of the two criteria presented in Equation~\ref{Bond criteria} is not verified. 

\begin{align}
    \overline{F_{si}}&\leq \mu \; \overline{F_{ni}} + \sigma_s A_b \text{ (Shear condition)}\nonumber\\
    \overline{F_{ni}}&\leq \sigma_n A_b \text{ (Tensile condition)}
    \label{Bond criteria}
\end{align}
where $\sigma_s$ is the shear strength of the bond, $\sigma_n$ is the tensile strength of the bond and $A_b$ is the surface of the bond. 
Similarly, $\mu$ is the friction coefficient between two particles, and it remains constant before and after the bond breakage.
It is worth noting that no criterion is employed for the compressive strength of the bond.


\subsection{Bond degradation model}

\begin{figure}[ht]
    \centering
    \includegraphics[width=0.7\linewidth]{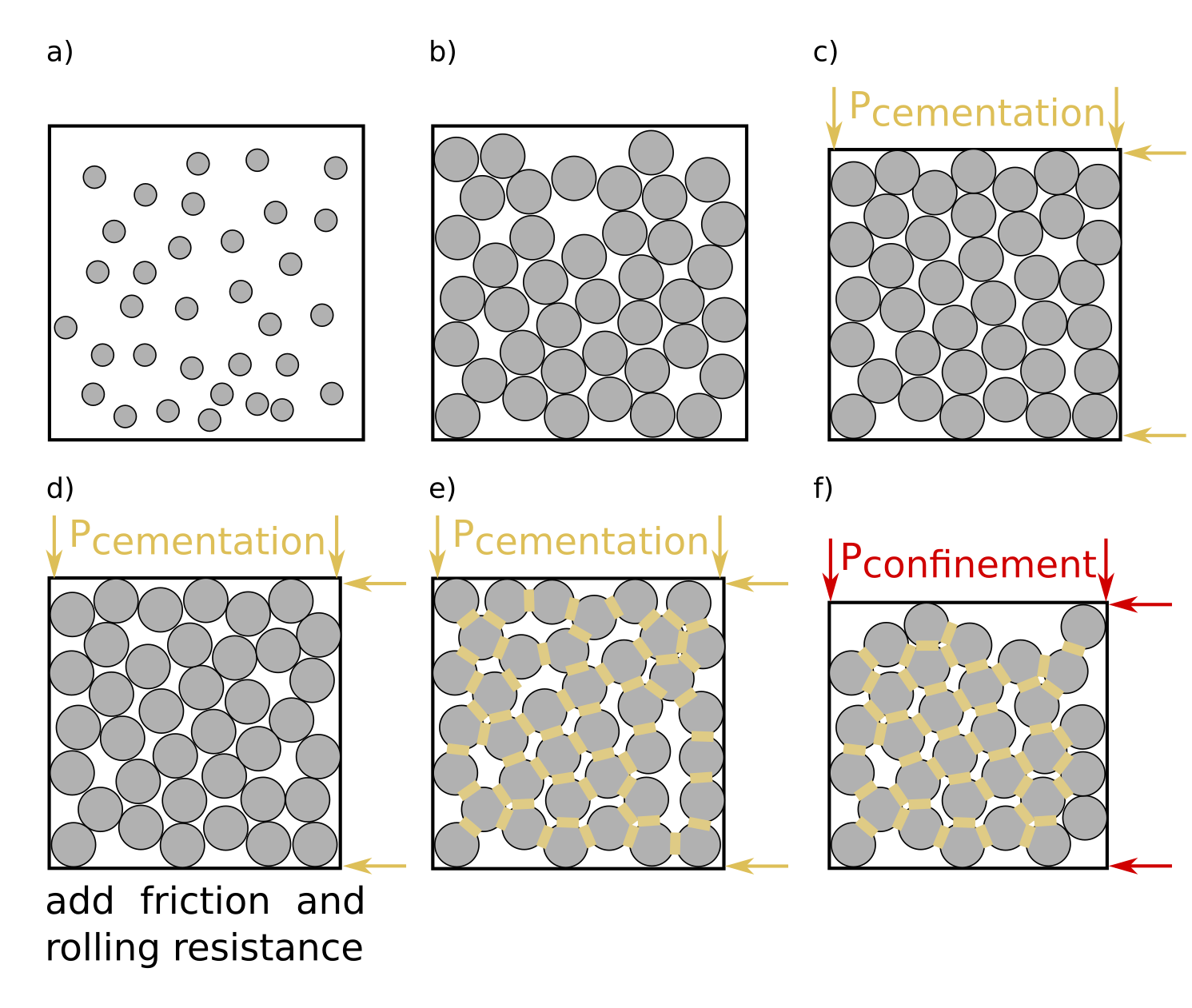}
    \caption{Initial condition algorithm: a) a box is created and all the particles are generated with a reduced radius. b) The different radii increase stepwise until the final values are reached. c) A confining pressure $P_{cementation}$ is applied. d) The friction, the twisting resistance, and the bending resistance are activated. e) The cementation is applied by generating bonds at the contacts. f) A confining pressure $P_{confinement}$ is applied, and the sample is ready to be characterized.}
    \label{Initial condition algorithm}
\end{figure}

To investigate the effect of the debonding, the algorithm presented in Figure \ref{Initial condition algorithm} and described in the following is applied to prepare the sample. 

A domain box is generated, the walls are considered to have no friction, cohesion, bending resistance, or twisting resistance with the grains.
Subsequently, particles are generated to obtain a Representative Elementary Volume. It is important to note that the grains are initially incorporated with a radius smaller than the final one. Subsequently, a radius expansion algorithm is applied to generate the initial condition \citep{OSullivan2011} and to verify the grain size distribution of the sample.

Once the sample is generated, the isotropic confinement is applied by controlling the position of the top+lateral walls in order to apply a confining pressure equal to $P_{cementation}$. 
Once the confinement reaches the requested value, the friction between the grains, the twisting resistance, and the bending resistance are activated. Similar to the previous step, the positions of the walls are controlled to verify the confinement $P_{cementation}$.

Subsequently, the cementation between the particles is applied. At each contact, a bond is formed with a probability $p_c$. 
The bonds are defined by a surface area $A_b$ obtained by a bond size distribution that characterizes the degree of cementation when the sample is intact.
This cementation generation also includes the application of the chemical damage.
Indeed, the surface area of the bond is herein reduced by a quantity $\xi_s$.
The effective surface area of the bond is $A_b^{eff}= max \left(A_b - \xi_s, 0\right)$. If the effective surface area $A_b^{eff}$ reaches the value $0\,\mu m^2$, the bond is not generated and the contact is considered uncemented.
In the following, the local stiffness at each contact $E_m(A_b^{eff})$ is determined by the effective bond surface \citep{SacMorane:StressState,DeSimone2025,Sun2016,Sun2018}, see Equation \ref{Young Local}. 

\begin{equation}
    E_m(A_b^{eff}) = \frac{E_m^{cementation}-E_m^{untreated}}{A_{b,\,ref}^{cementation}}\times A_b^{eff} + E_m^{untreated}
    \label{Young Local}
\end{equation}
where $E_m$ is the Young modulus of the contact, $E_m^{cementation}$ is the initial Young modulus depending on the initial degree of cementation, $E_m^{untreated}$ is the Young modulus without cementation, $A_{b,\,ref}^{cementation}$ is a bond surface of reference depending on the initial degree of cementation, and $A_b^ {eff}$ is the effective bond surface of the contact (after the reduction of the surface).
The degree of cementation is represented through the parameters $E_m^{cementation}$ and $A_{b,\,ref}^{cementation}$. This bond surface of reference  $A_{b,\,ref}^{cementation}$ is computed to ensure that the mean Young modulus within the intact sample is equal to $E_m^{cementation}$ after the cementation step depicted in Figure \ref{Initial condition algorithm}e. 
Details on how to obtain this coefficient are available in Appendix B from \citep{Wu2025b}. 
A distribution of the Young modulus $E_m$ is then induced within the sample. 
It is fundamental to emphasize that the bonds are formed after an initial loading step.
This assumption implies that the microstructures are stable without the cement, in contrast to the metastable state with the cementation that exists regularly in nature.
Even if the sample can be loaded after the cementation to generate this metastable state \citep{SacMorane:StressState}, this investigation remains out of the context of this contribution.

Once the weathered sample is generated, the positions of the walls are controlled to ensure a confinement equal to $P_{confinement}$.
At the end of this stage, the initial configuration is reached, and the estimation of the elastic properties can start.


\subsection{Macroscopic property extraction}
\label{Section Macro Prop Extract DEM}

To estimate its properties, the sample is loaded at a given bond surface reduction, see Algorithm \ref{Scheme Properties Estimation}.

\begin{algorithm}[ht]
\caption{The softening law induced by the chemical damage of the cemented granular material is approximated by decreasing the bond size stepwise.}
\label{Scheme Properties Estimation}
\begin{algorithmic}
\ForAll {surface reduction $\xi_s \in$ inputs}
    \State prepare the initial conditions \Comment{see Figure \ref{Initial condition algorithm}} 
    \State add small perturbations in the loading conditions
    \While{DEM equilibrium not reached} \Comment{granular reorganization occurs} 
        \State compute applied forces         \State solve the momentum balances
    \EndWhile
    \State estimate the sample parameters $\Pi$ \Comment{at a given $\xi_s$}
\EndFor
\State approximate the relation $\Pi(\xi_s)$
\end{algorithmic}
\end{algorithm}

In particular, two loading conditions are employed: the isotropic loading (Section \ref{Isotropic Loading}) and the triaxial loading (Section \ref{Triaxial Loading}). 
The details of the numerical tests are provided in the following Subsections.
The main idea is to consider a small perturbation in the loading conditions $\left(P_{load}^{i+1}=P_{load}^{i}+dP_{load}\right)$ to determine effective sample parameters, such as bulk modulus, shear modulus, Young modulus, and Poisson ratio.
Similar to methods available in the literature \citep{Magnanimo2008, Gong2019, Cheng2020, Reddy2022}, this perturbation is applied incrementally to ensure the quasi-static conditions.
Changing the loading conditions, the reorganization of the grains occurs until an equilibrium state is reached. Two indices determine this mechanical equilibrium:
\begin{itemize}
    \item the unbalanced force (defined as the ratio of the mean summary force on bodies and mean force magnitude on interactions) is smaller than a criterion value (here $0.01$).
    \item the difference between the pressures applied on the controlled walls and the targeted pressure is smaller than a criterion value (here $0.01\,P_{target}$). 
\end{itemize}

Once the equilibrium is reached, a snapshot of the sample is taken. This operation enables the tracking of the evolution of the various parameters, see the results of these numerical homogeneizations in Section \ref{Results Section DEM} for the discrete approach.

\subsubsection{Isotropic loading}
\label{Isotropic Loading}

The first elastic parameter to be determined is the bulk modulus $K$ through an isotropic loading test.
The top and the lateral walls are controlled to apply an isotropic loading $P_{load}$. 
Then, a small perturbation $dP_{load}$ is considered in the pressure applied at the top and lateral walls to induce an isotropic compression of the sample. 
Subsequently, the bulk modulus $K$ can be estimated from the relation \ref{Bulk Interpolation}.
It is worth specifying that the modulus estimated is the tangent property.
Furthermore, as no bond breakage occurs, this tangent property is also the secant property at a given state of chemical damage. 

\begin{equation}
    dP_{load} = K\times d\epsilon_v
    \label{Bulk Interpolation}
\end{equation}
where $d\epsilon_v=d\epsilon_x+d\epsilon_y+d\epsilon_z$ is the increment of the volumetric strain ($\epsilon_i$ is the strain in the direction $i$).

\subsubsection{Triaxial loading}
\label{Triaxial Loading}

Subsequently, the other elastic parameters (shear modulus $G$, Young modulus $E$, and the Poisson ratio $\nu$) are determined from a triaxial loading test.
The top wall displacement is controlled to apply a vertical loading $P_{load}$, and the lateral wall displacement is controlled to apply a confining loading $P_{confinement}$. 
Then, a small perturbation $dP_{load}$ is considered in the pressure applied at the top, while the positions of the lateral walls are controlled to apply the constant confining pressure $P_{confinement}$. 
The main consequence of the loading is a vertical settlement and a lateral expansion of the sample. 
Subsequently, the shear modulus $G$ can be estimated from the relation \ref{Shear Interpolation} \citep{Reddy2022}, the Young modulus $E$ can be estimated from the relation \ref{Young Interpolation}, and the Poisson ratio $\nu$ can be estimated from the relation \ref{Poisson Interpolation}.

\begin{equation}
    dq = 3\times G\times d\epsilon_q
    \label{Shear Interpolation}
\end{equation}
where $dq=P_{load}+dP_{load}-P_{confinement}$ is the increment of the deviatoric stress, and $d\epsilon_q=2/3\times\left(0.5(\epsilon_z-\epsilon_x) + 0.5(\epsilon_z-\epsilon_y)\right)$ is the increment of the deviatoric strain ($\epsilon_i$ is the strain in the direction $i$).

\begin{equation}
    dP_{load} = E\times d\epsilon_z
    \label{Young Interpolation}
\end{equation}
where $d\epsilon_z$ is the increment of the vertical strain.

\begin{equation}
    0.5(d\epsilon_x + d\epsilon_y) = -\nu\, d\epsilon_z
    \label{Poisson Interpolation}
\end{equation}
where $\epsilon_i$ is the strain in the direction $i$.

\section{The continuous approach}
\label{PF+FFT Section}

Even if DEM is applied in Section \ref{DEM Section} to predict the mechanical behavior of a cemented granular material during debonding, it appears that the geometry of the microstructure is simplified.
For instance, the cement phase located between the particles of the granular skeleton is not explicitly described.
And, the experimental observation of cemented granular material with X-ray tomography (XRCT) scans reveals that the geometry of the bond is far from being a cylinder joining only two grains (assumption made with the DEM).
Indeed, the cement material appears as irregular geometries and may form clusters of multiple particles \citep{Dadda2017, Tengattini2023, Tengattini2023b}.

To estimate the influence of these heterogeneities on the sample behavior, it becomes pivotal to explicitly describe the cement material by considering the rock as a material composed of three distinct phases (grain, cement, and pore spaces). In particular, these phases can be represented in the PF formulation by a combination of phase variables (one phase for the grain, one for the cement, and the pore space is determined by complementarity).
Subsequently, the PF method, specifically the Allen-Cahn equation presented in Section \ref{Theory PF}, is employed to describe the dissolution of the material \citep{Allen1979, Xu2008, Moelans2008, Takaki2014, Yang2021}.
In this approach, nonconserved order parameters are used to represent the mass dissolution.
Other formulations exist for the PF description of dissolution \citep{Lee2016, Wu2017, Aihara2019, Rohde2021, Li2023}, but they face additional challenges in being incorporated into a numerical model.

\begin{figure}[h]
    \centering
    \includegraphics[width=0.8\linewidth]{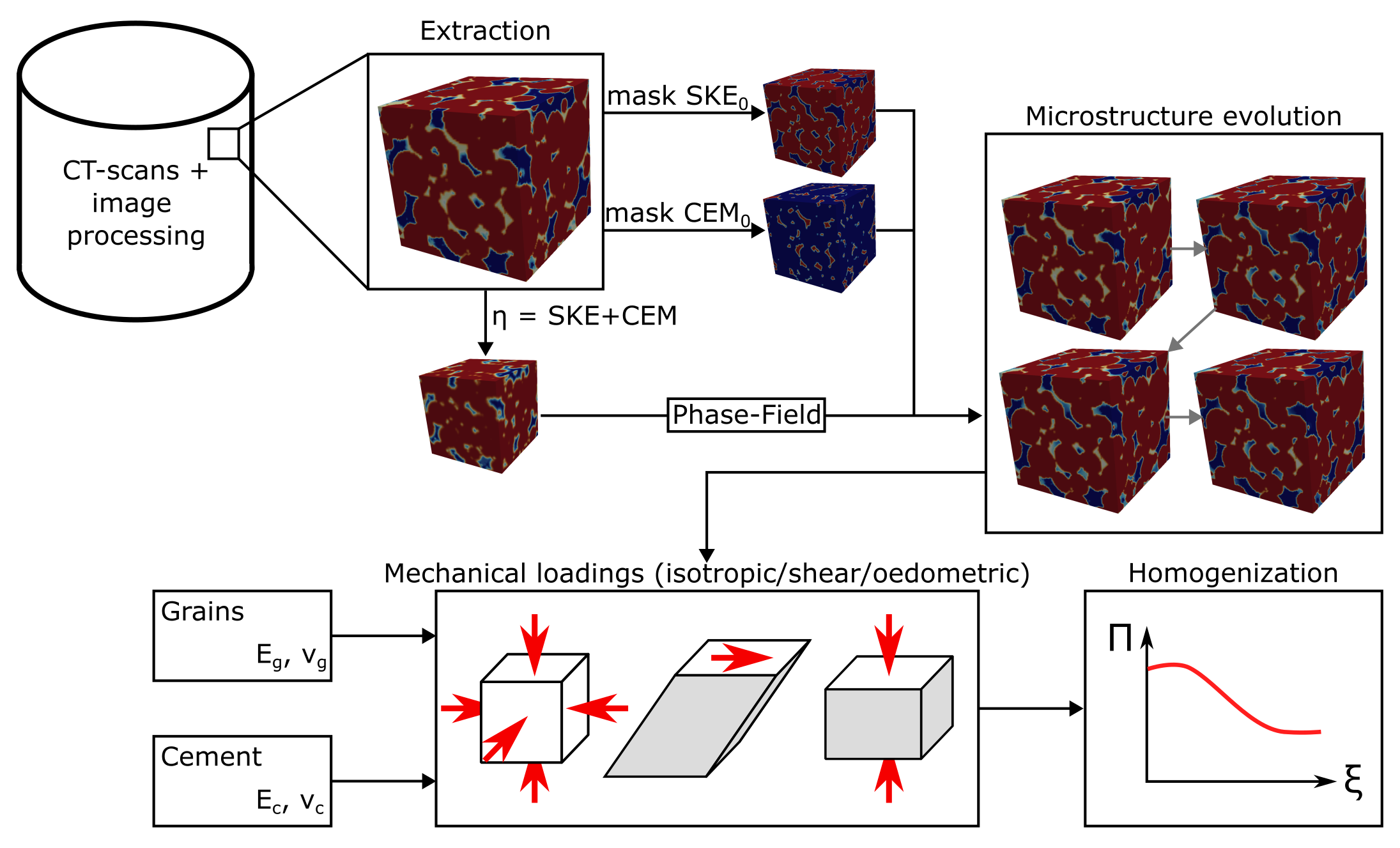}
    \caption{Diagram explaining the numerical homogenization scheme employed to compute the mechanical parameters of the Representative Elementary Volume from the prediction of the evolution of its microstructure.}
    \label{Homogenization Scheme}
\end{figure}

Once the PF description predicted the microstructure evolution of the material, mechanical loading is applied to determine the homogenized parameters, see Figure \ref{Homogenization Scheme}.
As depicted in Figures \ref{weathering rock}b and \ref{Elastic Parameters vs Chemical Damage}, the bond material quantity $V_b^{eff}$ is stepwise reduced by $\xi_v$ to investigate the impact of the debonding on the elastic properties of the material. 
Such a method has already been applied to predict, for instance, the mechanical performance of weathered sandstone \citep{Qiao2026} or of cement-based material during the hydration process \citep{SacMorane:Cementation, Nguyen2024}.
It is worth noting that the chemical aspect of the problem (microstructure evolution) is paused during this mechanical estimation of the composite material. 
Similar to the method presented in Section \ref{DEM Section}, the distinct microstructures are employed as independent inputs in the following. 
Furthermore, the impact of the mechanics on the chemistry and microstructure evolution is neglected herein. However, this feedback influence is pivotal in some contexts, such as the pressure-solution \citep{Weyl1959,Rutter1976,Tada1989}, and can be considered by advanced numerical methods.
For instance, a stress-dependent microstructure evolution has already been proposed in \citep{Guevel2020}. In a similar fashion, \citep{SacMorane:PFDEM, SacMorane:PFDEMb} depict the development of a new method to consider the impact of the stress and the microstructure reorganization on the chemistry.

While the numerical homogenization employed in the following would be possible with a Finite Element Model (FEM) \citep{SacMorane:Cementation, Nguyen2024}, a FFT formulation is preferred herein due to time efficient calculation \citep{Alavoine2020, Alavoine2020b,OlarteGarzon2027}. For instance, the time for an individual estimation of an element with 3.375.000 DOFs with the FFT method was 1 minute on a computer with Intel Core i5 vPro and 32 Gb of memory. 
For comparison, the application of the FEM appeared limited by the computational cost required (insufficient memory and long duration of the simulations for a computer with the preceding specifications).



\subsection{PF-based microstructure predictions}
\label{Theory PF}

The prediction of the microstructure evolution during the debonding is conducted with a PF formulation.
In particular, the Allen-Cahn equation \citep{Allen1979} given in Equation \ref{AC Equation} has been widely applied in the literature to solve the dissolution of a material.

\begin{equation}    
    \frac{\partial\, \eta}{\partial t}=-L \frac{\left(\partial f_{loc} + E_d\right)}{\partial\, \eta}+L\, \kappa\, \nabla^2 {\eta}
    \label{AC Equation}
\end{equation}
where $\eta$ is a phase variable to model the material, equaling 1 within the material and 0 elsewhere. 
Furthermore, $L\;(=1 \text{ herein})$ is the order parameter mobility, which affects the overall dissolution rate.
Similarly, $\kappa$ is the gradient energy coefficient, and it impacts the interface width and the reaction kinetics. This coefficient is related to the barrier energy $W$ and the mesh size $\Delta x$ $\left(\kappa\propto W\cdot\Delta x^2\right)$ \citep{SacMorane:PFDEMb,Takaki2014}.
The term $E_d$ encompasses additional energy sources introduced into the system (origin of the dissolution), such as mechanical, chemical, or thermal loading. In the system described herein, the destabilization energy is due to the chemical reaction between the cement and the pore fluid.
Finally, the local free energy density $f_{loc}$ is specified in Equation \ref{free energy Equation}.

\begin{equation}
    f_{loc}({\eta}) = W\times{\eta}^2(1-{\eta})^2 
    \label{free energy Equation}
\end{equation}

The local free energy density $f_{loc}$ is a double-well function with a barrier height $W\;(=1 \text{ herein})$ \citep{Takaki2014}. 
It is worth noting that the minima of this potential energy are located at $\eta= 0 \text{ and } 1$.
At the equilibrium and without external destabilization $(E_d=0)$, the phase variable stays at $\eta= 0 \text{ or } 1$ (no dissolution).
To obtain localized dissolution, an external source term $E_d$ is added to tilt the initial free energy, as depicted in Equation \ref{Ed Equation}.
Herein, the double-well function is tilted toward $\eta=0$ through the application of the amplitude $e_d$ ($=0.4\, W$ herein), which affects the kinetics of the dissolution.

\begin{equation}
    E_d(\eta) = e_d\times \left(3\,{\eta}^2-2\,{\eta}^3\right)    
    \label{Ed Equation}
\end{equation}

\vskip\baselineskip

As depicted in Figure \ref{Homogenization Scheme} and explained above, the microstructure of a cemented granular material can be represented by a combination of two phase variables \emph{SKE} (the granular skeleton) and \emph{CEM} (the cement).
Only the cement phase is affected by the debonding phenomenon.
However, the grain phase has a primordial role in the dissolution pattern of the cement.
For instance, the grains will prevent the dissolution of the cement if this phase is trapped in it; the pore fluid can not reach the reactive surface of the cement.
To consider this effect, a new variable is defined $\eta=$ \emph{SKE}$\,+\,$\emph{CEM}, and two masks are saved at the initial configuration (no chemical damage) \emph{SKE}$_0$ and  \emph{CEM}$_0$, see Figure \ref{Homogenization Scheme}.
The Allen-Cahn formulation presented in Equation \ref{AC Equation} is then applied to the variable $\eta$.
Subsequently, the microstructure is reconstructed by the relations  \emph{SKE}$\,=\,$\emph{SKE}$_0$ and \emph{CEM}$\,=\eta\,\cdot\,$\emph{CEM}$_0$.
By doing so, it is possible to identify the grains (remaining unaffected by the dissolution) and the cement phase. 
The duration of the simulation corresponds to the chemical damage applied to the sample.
Compared to Section \ref{DEM Section} where the bonds are modeled with a surface, the approach described herein considers the cement in an explicit manner, enabling the measurement of its volume.
Then, a new variable $\xi_v$ is introduced for the bond reduction as $\xi_v=\left(V_b^0 - V_b^{eff}\right)/V_{sample}$, where $V_b^0$ and $V_b^{eff}$ are the initial and current volumes of cement and $V_{sample}$ is the volume of the sample.
This variable is directly related to the mass removal of the cement.
An auxiliary debonding variable $\hat{\xi}=\xi_v/\xi_{v,\,max}$ ($\xi_{v,\,max}$ is the volume of cement in the intact state) is also introduced to translate the advancement of the weathering phenomenon ($=0$ for an intact sample and $=1$ for the state with a complete cement dissolution).
To determine the evolution of the elastic parameters of the sample while debonding occurs, an individual PF simulation is conducted, and several snapshots are extracted, following the scheme depicted in Figure \ref{Homogenization Scheme}.

\subsection{FFT-based homogeneization}
\label{Theory FFT}

Once the microstructure evolution has been predicted, the numerical homogenization scheme tackles the estimation of the elastic properties at the sample scale.
Continuous micromechanics is a theoretical framework for scaling the elastic properties of composite materials based on microstructural information. Pioneering work proposed by \emph{Eshelby} \citep{eshelby1957determination,eshelby1959elastic} on the concentration of elastic fields in an ellipsoidal inhomogeneity laid the foundations for so-called homogenization schemes. Introduced by \emph{Moulinec and Suquet} \citep{moulinec1994fast, moulinec1998numerical}, FFT-based methods have become strong competitors to standard numerical approaches \citep{Schneider2021, Lucarini2022}. The use of FFT provides an efficient approach for solving micromechanical problems, which involve applying the principles of mathematics, mechanics, and physics to describe the behavior of geomaterials \citep{Alavoine2020,Alavoine2020b,OlarteGarzon2027}.

The method represents a heterogeneous material through a periodic unit cell discretized into a regular grid of material points with spatially varying properties. Microscopic fields are mapped onto this grid, and the FFT is employed to solve a single integral equation. The mechanical problem is defined for a periodic domain $\Omega$ and solved locally. Indeed, this domain is subjected to a macroscopic deformation load $\Strain$ applied at the scale of the sample, and the resulting displacements locally fluctuate about their macroscopic counterpart $\Strain \cdot \x$, where $x$ is the local space variable in this domain. In periodic homogenization, the microstructure as well as the fluctuations of the displacement $\tilde{\tenone{u}}(\x)$ are periodic, so the local problem can be read as:

\begin{equation}
\label{eq:local_problem}
	\left\{
	\begin{array}{ll}
		\div{\stress(\x)} = \tenone{0}                         & \\
		\stress(\x) = \CC(\x) : \strain(\x)  & \\
		\strain(\x) = \tenone{\nabla}^{S}(\tenone{u}(\x))                      & \\
		\tilde{\tenone{u}}(\x)  \quad periodic                       & \\
		\stress \cdot \tenone{n} \quad antiperiodic     &
	\end{array}
	\right.
	\quad
	\begin{array}{ll}
		(\Omega)                     & \\
		(\Omega)                     & \\
		(\Omega)                     & \\
		(\delta \Omega)  & \\
		(\delta \Omega) &
	\end{array}
\end{equation}

\noindent where $\stress(\x)$ is the stress tensor, $\CC(\x)$ is the fourth-order stiffness tensor, $\strain(\x)$ is the strain tensor, and $\tenone{\nabla}^{S}(\tenone{u}(\x))$ is the symmetric gradient of the displacement. The first three lines represent respectively the equilibrium equations, the constitutive equations, the compatibility equations, and the last two lines the periodic boundary conditions. \emph{Moulinec and Suquet} \citep{moulinec1994fast, moulinec1998numerical} solved the local mechanical problem by introducing a homogeneous reference material of arbitrary properties $\CC^0$. This reference medium is used to quantify the local strain fluctuations induced by elastic heterogeneities, which are measured relative to the strain field
of the homogeneous case. The result is an implicit integral equation known as the Lippmann-Schwinger equation which is defined by Green’s function $\tenfour{\Gamma}^0$ and reads as:

\begin{equation}
	\label{Lippmann-Schwinger}
    \strain(\x) = \Strain - \tenfour{\Gamma}^0 \star \left[(\CC(\x) - \CC^0) : \strain(\x)\right]
\end{equation}

\noindent with $\star$ as the convolution product. In periodic elasticity, the convolution product in Equation \ref{Lippmann-Schwinger} is most conveniently evaluated in Fourier space,
where the operator $\tenfour{\Gamma}^0$ is explicitly known. 
The reader is referred to the recent review \citep{Schneider2021} for a more detailed description of this numerical technique. Regardless of spatial discretization, the resulting Lippmann-Schwinger equation cannot be solved in closed form and requires an iterative
algorithm \citep{willot2015fourier}. 
This method follows the basic scheme proposed by \emph{Moulinec and Suquet} \citep{moulinec1994fast, moulinec1998numerical}, which solves the Lippmann-Schwinger equation through a fixed-point iteration equivalent to a Neumann series.
It is important to specify that the number of iterations can become excessive for large or infinite contrasts. Thus, some authors have focused on developing algorithms that compare to the original method in terms of convergence rate or the use of infinite contrast \citep{eyre1999fast, monchiet2012polarization, schneider2017fft,schneider2019barzilai, sab2024fft, dolbeau2024accelerating}. The so-called adaptive Eyre–Milton (AEM) scheme proposed by \emph{Sab et al.} \citep{sab2024fft} is employed in the following. This iterative scheme remains meaningful even in the notoriously difficult case of heterogeneous materials that contain both pores and rigid inclusions.
A more detailed formulation of the solver employed is available in \citep{OlarteGarzon2027}.


\subsection{Macroscopic property extraction}

The sample is loaded by the perturbation method at a state of given mass removal to estimate its effective mechanical properties.
In particular, three loading conditions are employed: the isotropic loading (Section \ref{Isotropic Loading 2}), the oedometric loading (Section \ref{Oedometric Loading}), and the simple shear loading (Section \ref{Shearing Loading}).
The main idea is to consider a small perturbation in the loading conditions to determine effective parameters of the material, such as bulk modulus, shear modulus, Young modulus, and Poisson ratio.
This operation enables tracking the evolution of the various properties, see the results of this numerical homogenization in Section \ref{Results Section PF} for the continuous approach.

\subsubsection{Isotropic loading}
\label{Isotropic Loading 2}

In the same note as Section \ref{Isotropic Loading}, the bulk modulus $K$ is estimated through an isotropic loading test.
The only difference is the fact that the loading is herein controlled by the displacement and not the stress, applying an incremental isotropic compression of the sample with $d\Strain_{xx}=d\Strain_{yy}=d\Strain_{zz}$.
The consequence of this mechanical solicitation is the increase of the mean stress $d\Stress_v=\left(d\Stress_{xx}+d\Stress_{yy}+d\Stress_{zz}\right)/3$ in the material.
Subsequently, the bulk modulus $K$ can be estimated from the relation \ref{Bulk Interpolation} (considering $d\Stress_v$ for $dP_{load}$ and $d\Strain_v=d\Strain_{xx}+d\Strain_{yy}+d\Strain_{zz}$ for $d\epsilon_v$).

\subsubsection{Simple shear loading}
\label{Shearing Loading}

Then, the shear modulus is obtained from a simple shear loading condition.
A strain increment $d\Strain_{xz}$ is applied.
The main consequence of this perturbation is the incremental increase in the shear stress transmitted inside the sample, which can be approximated by its average value $d\Stress_{xz}$. 
The shear modulus $G$ can be estimated from the relation \ref{Shear Interpolation 2}.

\begin{equation}
    d\Stress_{xz} = G\times 2\,d\Strain_{xz}
    \label{Shear Interpolation 2}
\end{equation}

\subsubsection{Oedometric loading}
\label{Oedometric Loading}

The Young modulus is determined from an oedometric loading condition as the FFT-based solver employed herein only implements strain-controlled boundary conditions.
The macroscopic loading consists of an incremental uniaxial strain $d\Strain_{zz}$ while the lateral deformations are prevented $\left(d\Strain_{xx} = d\Strain_{yy} = 0.0\right)$.
The Young modulus oedometric $E_{oedo}$ can be estimated from the relation \ref{Young Oedometric Interpolation}.

\begin{equation}
    d\Stress_{zz} = E_{oedo}\times d\Strain_{zz}
    \label{Young Oedometric Interpolation}
\end{equation}
where $d\Stress_{zz}$ is the average value of the vertical component of the incremental stress due to the mechanical deformation.

It is worth noting that the modulus obtained with an oedometric test is different than the Young modulus of the material.
Indeed, the lateral conditions of the oedometric loading, preventing the lateral displacement, contain and reinforce the sample.
This modulus should be adapted to determine the real Young modulus of the sample.
To do so, the $k_0$ coefficient is determined through the relation $k_0=dP_{lat}\,/\,dP_{load}$, where $dP_{lat}$ is the average increment for the normal stress on the lateral faces and $dP_{load}$ on the vertical faces. Then, the Poisson ratio $\nu$ can be determined through the relation \ref{Poisson Oedometric Interpolation}.

\begin{equation}
    \nu = \frac{k_0}{1+k_0}=\frac{dP_{lat}}{dP_{load}+dP_{lat}}
    \label{Poisson Oedometric Interpolation}
\end{equation}

Finally, the Young modulus of the sample is determined through the relation \ref{Young 2 Interpolation}.

\begin{equation}
    E = E_{oedo}\frac{(1+\nu)(1-2\nu)}{1-\nu}
    \label{Young 2 Interpolation}
\end{equation}


\section{Numerical applications and results}
\label{Results Section}

\subsection{Discrete element approach}
\label{Results Section DEM}

In this Section, a calibrated model of biocemented sand \citep{Sarkis2022, Dadda2017} is employed to represent natural sandstones.
The DEM is solved using the \emph{YADE} open source software \citep{YADE}. 
Some examples of scripts used for this contribution are available on GitHub \citep{LinkGithub}. 
Herein, $3000$ particles are generated, despite $1600$ would suffice to obtain a Representative Elementary Volume \citep{Sarkis2022,OSullivan2011}. These grains verify the uniform grain size distribution ($R_{min}=75$~$\mu$m and $R_{max}=125$~$\mu$m) that characterizes the samples used for the calibration conducted by \citep{Sarkis2022}.

During the cementation, step e) of Figure \ref{Initial condition algorithm}, the bonds are defined by a surface area $A_b$ obtained by a lognormal distribution presented in Appendix \ref{Lognormal Distribution} and defined by the parameters $m_{log}$ and $s_{log}$, respecting the observations from \citep{Sarkis2022}. 
The parameters $p_c$, $m_{log}$, and $s_{log}$ characterize the degree of cementation when the sample is intact.
In particular, this paper investigates four distinct initial cementation levels, see the parameters summarized in Table \ref{Parameters DEM}.  
The samples (2MB, 11BB, 13BT, 13MB) are referenced by the same nomenclature as in \citep{Sarkis2022} based on the sample preparation \citep{Dadda2017}, where the number signifies the column index and the letters stand for the location of the sample in the column (MB=middle bottom, BB=bottom bottom, BT=bottom top).
Even if the sample name is not related to the degree of cementation, the order 2MB < 11BB < 13BT < 13MB is respected in terms of cement quantity.

\begin{table}[h]
    \caption{Mechanical parameters depending on the cementation (extracted from \citep{Sarkis2022}).}
    \centering
    \begin{tabular}{|p{0.2\linewidth}|c|c|c|c|c|}
        \multicolumn{2}{c}{}&\multicolumn{1}{|p{0.13\linewidth}|}{Lightly cemented}&\multicolumn{1}{p{0.13\linewidth}}{Medium cemented}&\multicolumn{2}{|p{0.13\linewidth}|}{Highly cemented}\\
        \hline
        Sample&Untreated&2MB&11BB&13BT&13MB\\
        \hline
        Density (kg/m$^3$)&\multicolumn{5}{c|}{$2650$}\\
        \hline
        $E_m$ (MPa)&$80$&$320$&$760$&$860$&$1000$\\
        \hline
        $\nu_c$ (-)&\multicolumn{5}{c|}{$0.25$}\\
        \hline
        $\alpha_b$ (-)&\multicolumn{5}{c|}{$0.5$}\\
        \hline
        $\alpha_t$ (-)&\multicolumn{5}{c|}{$0.5$}\\
        \hline
        $\mu$ (-)&\multicolumn{5}{c|}{$0.36$}\\        
        \hline
        $p_c$ (\%)&$0$&$88$&$98$&$100$&$100$\\  
        \hline
        $m_{log}$ (-)&$0$&$7.69$&$8.01$&$8.44$&$8.77$\\
        \hline
        $s_{log}$ (-)&$0$&$0.60$&$0.88$&$0.92$&$0.73$\\   
        \hline
        $A_{b,\,ref}^{cementation}\; (\mu m^2)$& $\infty$ & $2303$ & $4267$ & $6577$ & $8131$\\
        \hline
        $\sigma_s$ (MPa)&\multicolumn{5}{c|}{$10\times6.6$}\\ 
        \hline
        $\sigma_n$ (MPa)&\multicolumn{5}{c|}{$10\times2.75$}\\ 
        \hline
    \end{tabular}
    \label{Parameters DEM}
\end{table}

A parameter adjustment was applied compared to the values calibrated in \citep{Sarkis2022}: the strength of the bonds is artificially increased by a factor $10$, in order to prevent bond breakage during the loading tests and simulate softening of the elastic parameters due to chemical alterations solely.
This artificial modification of the bond strength may introduce non-physical behavior at the mechanical rupture of the material, however, it would not affect the reduction of the elastic properties due to chemical solicitations, which are the focus of this investigation. 
A complementary discussion on this no bond breakage is available in Section \ref{Use Discussion Section}.

For the numerical values employed in the mechanical loadings, this application considers the vertical loading $P_{load}$, the confinement pressure $P_{confinement}$, and the cementation pressure $P_{cementation}$ equal to $0.1$ MPa. The perturbation $dP_{load}$ to determine the elastic properties is $0.1\times P_{load}=0.01$ MPa.

The different configurations are repeated at least three times to ensure the reproducibility of the simulations. 
Moreover, the reference positions considered for the distinct strains are the positions after the initialization algorithm depicted in Figure \ref{Initial condition algorithm}f: $\epsilon_i=\Delta_i/L_i^0$, where $\epsilon_i$ is the strain in the $i$ direction, $\Delta_i$ is the dimension change in the $i$ direction, and $L_i^0$ is the dimension of reference in the $i$ direction.
It is worth specifying that this dimension of reference $L_i^0$ is not constant for all the simulations due to the sample generation algorithm depicted in Figure \ref{Initial condition algorithm}. 
However, this variation remain small, see the details in Appendix \ref{ev0}.
Concerning the estimation of the modulus, although some fluctuations occur, the individual $R^2$ metric computed is at least equal to 0.99996.

\vskip\baselineskip

\begin{figure}[h]
    \centering
    \includegraphics[width=0.9\linewidth]{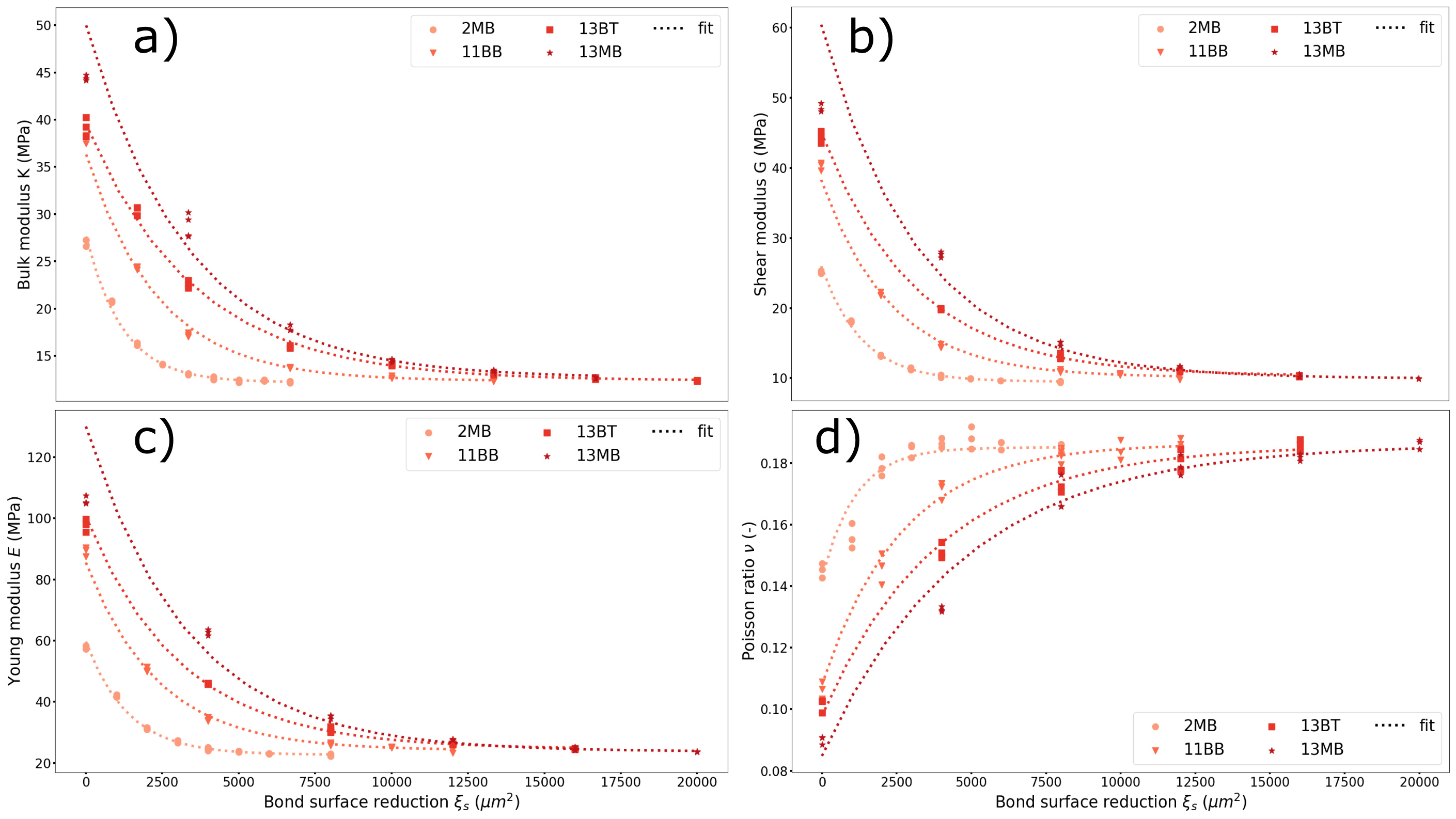}  
    \caption{Evolution of a) the bulk modulus, b) the shear modulus, c) the Young modulus, and d) the Poisson ratio with the bond surface reduction $\xi_s$ for distinct initial cementations.}
    \label{Results Figure Pi}
\end{figure}

The isotropic loading test depicted in Subsection \ref{Isotropic Loading} is employed to determine the bulk modulus by applying Equation \ref{Bulk Interpolation} to the sample behavior.
In the same way, the triaxial loading test depicted in Subsection \ref{Triaxial Loading} is employed to estimate the shear modulus by applying Equation \ref{Shear Interpolation} to the sample behavior.
Subsequently, the Young modulus is determined with Equation \ref{Young Interpolation} and the Poisson ratio is measured with Equation \ref{Poisson Interpolation}.
The evolution of these parameters is presented in Figure \ref{Results Figure Pi}.

As expected, the elastic properties decrease (increase for the Poisson ratio) with the surface reduction, and this reduction appears to follow an exponential-like decay. 
A generic variable $\Pi$ is used in the following to represent the effective sample properties $K$ (bulk modulus), $G$ (shear modulus), $E$ (Young modulus), or $\nu$ (Poisson ratio).
We introduce the expression $\Pi(\xi_s)=\left(\Pi_0-\Pi_\infty\right)\times \text{exp}(-b_\Pi\times \xi_s)+\Pi_\infty$, which fits accurately the simulation results, see Figure \ref{Results Figure Pi}.

\begin{table}[h]
    \caption{Parameters estimated for the relation $\Pi(\xi_s)=\left(\Pi_0-\Pi_\infty\right)\times exp(-b_\Pi\times \xi_s)+\Pi_\infty$ considering distinct cementations.}
    \centering
    \begin{tabular}{|l||c|c|c||c|}
        \hline
        \multicolumn{5}{|c|}{Bulk modulus $K(\xi_s)$}\\
        \hline
        cementation & $K_0-K_\infty$ (MPa) & $b_K$ ($\mu m^{-2}$)& $K_\infty$ (MPa) & $R^2$ \\
        \hline
        2MB  & 15.4 & 6.95$e^{-4}$ & 12.2 & 0.994\\
        11BB & 24.0 & 3.55$e^{-4}$ & 12.3 & 0.992\\
        13BT & 27.1 & 2.37$e^{-4}$ & 12.4 & 0.996\\
        13MB & 37.3 & 2.50$e^{-4}$ & 12.6 & 0.950\\
        \hline
        \multicolumn{5}{|c|}{Shear modulus $G(\xi_s)$}\\
        \hline
        cementation & $G_0-G_\infty$ (MPa) & $b_G$ ($\mu m^{-2}$)& $G_\infty$ (MPa) & $R^2$ \\
        \hline
        2MB  & 16.5 & 7.42$e^{-4}$ & 9.5 & 0.993\\
        11BB & 28.2 & 4.29$e^{-4}$ & 10.1 & 0.994\\
        13BT & 34.8 & 3.23$e^{-4}$ & 10.3 & 0.999\\
        13MB & 50.5 & 3.09$e^{-4}$ & 9.9  & 0.869\\
        \hline
        \multicolumn{5}{|c|}{Young modulus $E(\xi_s)$}\\
        \hline
        cementation & $E_0-E_\infty$ (MPa) & $b_E$ ($\mu m^{-2}$)& $E_\infty$ (MPa) & $R^2$ \\
        \hline
        2MB  & 37.0 & 7.31$e^{-4}$ & 22.7 & 0,985\\
        11BB & 61.4 & 4.24$e^{-4}$ & 24.1 & 0,993\\
        13BT & 76.0 & 3.22$e^{-4}$ & 25.0 & 0,995\\
        13MB & 106  & 3.01$e^{-4}$ & 24.0 & 0.958\\
        \hline
        \multicolumn{5}{|c|}{Poisson ratio $\nu(\xi_s)$}\\
        \hline
        cementation & $\nu_0-\nu_\infty$ (-) & $b_\nu$ ($\mu m^{-2}$)& $\nu_\infty$ (-) & $R^2$ \\
        \hline
        2MB  & -0.04 & 8.93$e^{-4}$ & 0.19 & 0.719\\
        11BB & -0.08 & 3.76$e^{-4}$ & 0.19 & 0.926\\
        13BT & -0.09 & 2.54$e^{-4}$ & 0.19 & 0,929\\
        13MB & -0.10 & 2.11$e^{-4}$ & 0.19 & 0.943\\
        \hline
    \end{tabular}
    \label{Results Table Pi}
\end{table}

The best-fit values of the coefficients $\Pi_0$, $b_\Pi$, and $\Pi_\infty$ were determined by least square error fitting and are specified in Table \ref{Results Table Pi}.
The coefficient $\Pi_0-\Pi_\infty$ represents the variation of the parameter between the intact and the unbonded states.
It depends on the mean Young modulus at the contact considered for the cementation $E_m^{cementation}$. 
As shown in Tables \ref{Parameters DEM} and \ref{Results Table Pi}, the absolute value of the coefficient $|\Pi_0-\Pi_\infty|$ increases with the contact Young modulus $E_m$.
Correspondingly, the coefficient $b_\Pi$ represents the softening rate of the parameter with the bond reduction.
This parameter depends on the bond size distribution of the sample, and in particular, on the mean cement size. Moreover, the coefficient increases with the uniformity of the bond size distribution. 
Similarly, this coefficient is also affected by the mean Young modulus at the contact, depicting the cementation degree. 
Indeed, at a given bond size distribution (which is not the case with the samples investigated herein), the parameter $b_\Pi$ should increase with respect to $E_m^{cementation}$ as $|\Pi_0-\Pi_\infty|$ becomes larger.
It is worth noting that the bond size distribution also defines the softening law of the parameter with the bond reduction. Depending on this distribution, the relation $\Pi(\xi_s)=\left(\Pi_0-\Pi_\infty\right)\times \text{exp}(-b_\Pi\times \xi_s)+\Pi_\infty$ may not be the best fit available and should not be applied blindly to represent the softening of a rock material during its alteration.

The observed dependency of the coefficient $G_0-G_\infty$ on the degree of cementation is in agreement with previous numerical investigations on bio-cemented soils \citep{Zhang2024}.

\vskip\baselineskip

Then, the quality of the approximated relations $\Pi(\xi_s)$ is characterized. The initial approach is to ensure that the unbonded state is similar for the distinct estimated relations. Indeed, this state remains the final and common state of the distinct cementations considered, and the value obtained for $\xi_s\rightarrow\infty$ of the elastic parameters must be similar even if the initial cementations are distinct.
It appears in Table \ref{Results Table Pi} that the elastic parameters for the determined unbonded states $\Pi_\infty$ appear consistent, with a maximum relative error of 5\% compared to the mean value. 
This error can be explained by a small quantity of remaining bonds (between 1\% and 8\%) that persisted after the maximum surface reduction applied in our simulations.
These residual bonds reinforce the sample compared to the unbonded state. A better estimation of the unbonded state could be done by increasing the bond surface reduction employed during the simulations. However, the quality of the results remains acceptable.

\begin{figure}[h]
    \centering
    \includegraphics[width=0.8\linewidth]{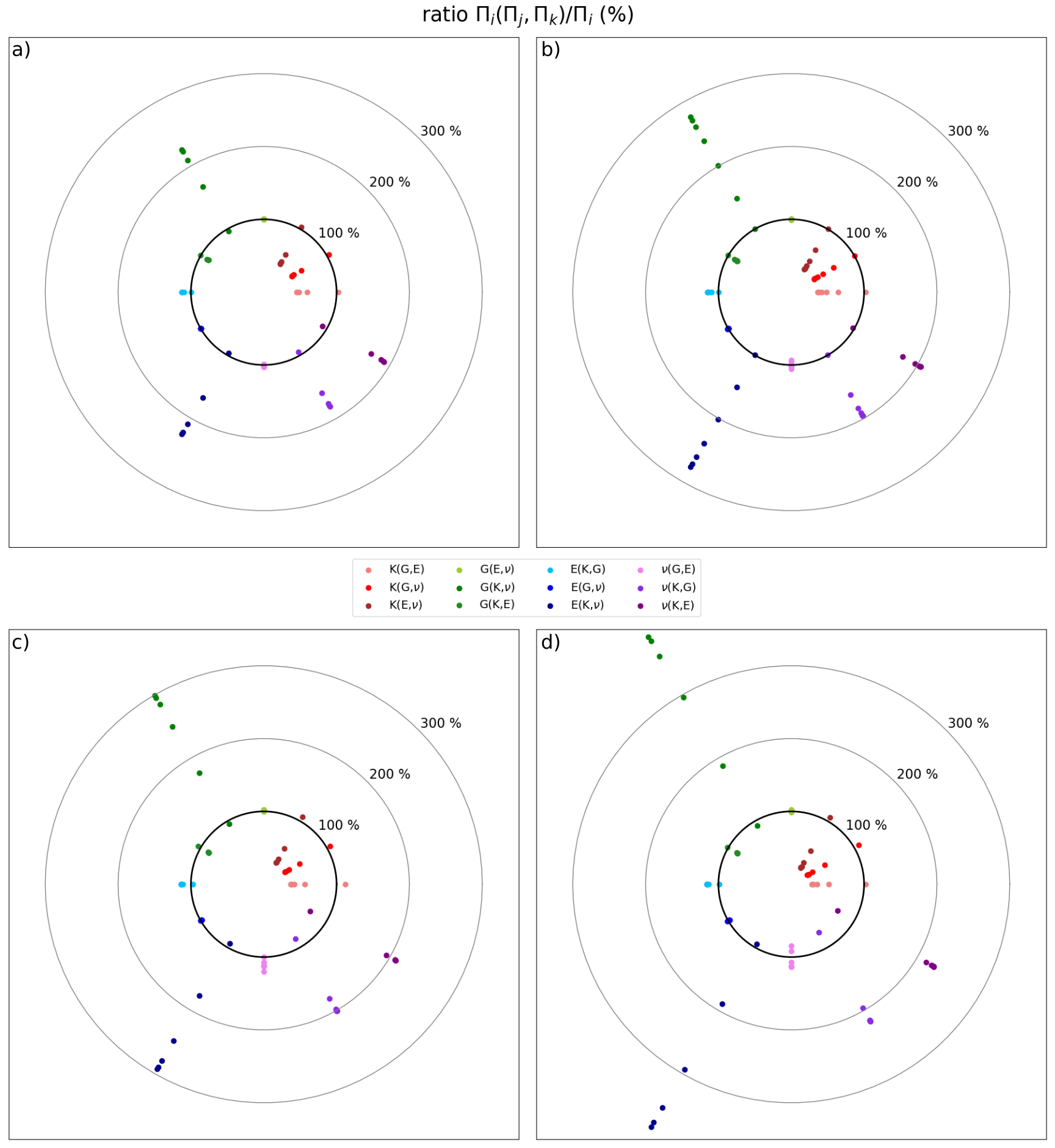}
    \caption{Cross verification of the interpolated parameters for an initial cementation a) 2MB, b) 11BB, c) 13BT, and d) 13MB.}
    \label{Results Figure Cross Verification}
\end{figure}

As recalled in Equations \ref{Equation K}-\ref{Equation v}, the distinct elastic parameters are correlated.
The goal is then to verify the parameter approximation with the relation $\Pi_i$ vs. $\Pi_i(\Pi_j,\,\Pi_k)$.    
In particular, Figure \ref{Results Figure Cross Verification} depicts the ratio between the approximated parameter and the estimated parameter $\Pi_i$ vs. $\Pi_i(\Pi_j,\,\Pi_k)$ for different surface reduction $\xi_s$. The different points must be located on the isoline ratio = 100\%.

\begin{align}
    &K   = \frac{E\cdot G}{3(3\cdot G-E)} = \frac{2\cdot G(1+\nu)}{3(1-2\cdot\nu)} = \frac{E}{3(1-2\cdot\nu)} \label{Equation K}\\
    &G   = \frac{E}{2(1+\nu)} = \frac{3\cdot K (1-2\cdot\nu)}{2(1+\nu)} =\frac{3\cdot K\cdot E}{9\cdot K- E} \label{Equation G}\\
    &E   = \frac{9\cdot K\cdot G}{3\cdot K+G} = 2\cdot G(1+\nu) =3\cdot K(1-2\cdot\nu) \label{Equation Y}\\
    &\nu = \frac{E}{2\cdot G}-1 = \frac{3\cdot K-2\cdot G}{2(3\cdot K + G)} = \frac{3\cdot K-E}{6\cdot K}  \label{Equation v}
\end{align}

It appears that the relations between the elastic parameters are verified as the points are located around the 100\% isoline.
However, some fluctuations can be observed for $K(E,\,G)$, $K(G,\,\nu)$, $K(E,\,\nu)$, $E(K,\,\nu)$, $G(K,\,\nu)$, $\nu(K,\,G)$, and $\nu(K,\,E)$.
Indeed, the approximation errors, due to the residual bonds and the fit relation, propagate. It is particularly the case if the distinct isotropic (used for $K$) and triaxial (used for $E,\, G,\, \nu$) tests are employed.
The ratios obtained considering $\xi_s=0$ are the most accurate as they are located on the 100\% isoline. The differences are accentuated while $\xi_s$ increases. 
It can be emphasized that if the triaxial test is employed only (see $E(G,\,\nu)$ and $G(E,\,\nu)$), the points are exactly located on the targeted 100\% isoline for all surface reduction $\xi_s$. 
It is less verified for $\nu(E,\,G)$ as the estimation is less accurate for the Poisson ratio, see the $R^2$ coefficients in Table \ref{Results Table Pi}.


\subsection{Continuous approach}
\label{Results Section PF}

In the philosophy of the Rocks Digital Physics \citep{Yang2026,Dvorkin2011,Andra2013,Lesueur2022}, which aims to determine the properties of a rock from its images obtained by XRCT-scans, an avatar of a real cemented sample is investigated.
Unfortunately, such data for the samples investigated in Section \ref{Results Section DEM} were unavailable.
Thus, the geometry of our numerical sample is taken from \emph{Tengattini et al.}, who studied spherical glass beads with calcite as a cement under triaxial loading \citep{Tengattini2023, Tengattini2023b}.
Even though this microstructure is simplified compared to natural cemented granular rock, which can exhibit irregular particles, it allows direct comparisons of behavior with the DEM spherical particle structure and with experimental data.

\begin{figure}[h]
    \centering
    \includegraphics[width=0.7\linewidth]{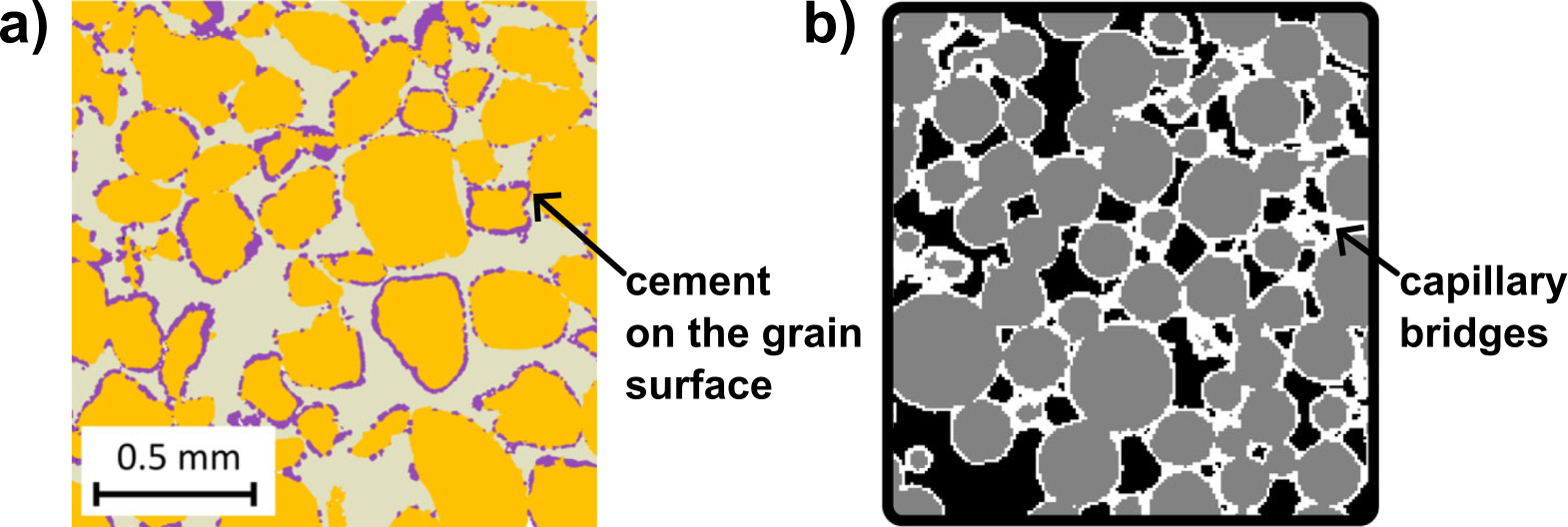}
    \caption{Cementation patterns for a) a biocemented sand \citep{Dadda2019} and b) cemented glass beads \citep{Tengattini2023}.}
    \label{Figure biocementation capillary bridge}
\end{figure}

It is essential to remind that the samples differ between Sections \ref{Results Section DEM} and \ref{Results Section PF} in terms of degree of cementation and bond and grain size distributions.
For instance, the cementation for the biocemented sands (Section \ref{Results Section DEM}) is located on the surface of the grains, while the cementation of the glass beads (current Section \ref{Results Section PF}) is based on capillary bridges.
As depicted in Figure \ref{Figure biocementation capillary bridge}, this fluctuation in the cementation mechanisms induces distinct patterns between the samples.
Indeed, the capillary bridges tend to create new contacts between the grains, while this generation remains limited for the biocementation. 
This fundamental difference in the microstructure may introduce distinct sample behavior during its alteration.

The size of the sample's Representative Elementary Volume is first determined (Appendix \ref{REV PF}), where a 
subset of 150x150x150 voxels seems as a good compromise between representativeness and computational cost.
Moreover, five subsets are considered to ensure the reproducibility of the simulations.

The prediction of the microstructure evolution is conducted by solving the Equation \ref{AC Equation} with the open-source FEM software MOOSE \citep{MOOSE, MOOSE_PF}.
Some examples of scripts used in this contribution are available on GitHub \citep{LinkGithubb}.
A movie example of the microstructure evolution is provided as Supplemental Material.

Concerning the estimation of the mechanical parameters, a Young modulus for the grains equal to $1$ Pa, see Table \ref{Parameters PF}, has been considered arbitrarily.
Following the approach from \citep{Lesueur2022}, the Young modulus for the cement is considered less rigid than the grains and different cement rigidities are explored to investigate the effect of this phase on the composite behavior, see Table \ref{Parameters PF}.
Finally, the pore phase is considered with no rigidity.
The Poisson ratio is assumed to be equal to $0.3$ for the two main phases.
As explained in Section \ref{DEM Section}, isotropic, oedometric, and shear loading conditions are applied to the sample, see the details in Sections \ref{Isotropic Loading 2}, \ref{Oedometric Loading}, and \ref{Shearing Loading}.
In particular, $d\Strain_{ii}=0.033$ is applied to obtain $d\Strain_v=0.1$ for the isotropic loading. 
By the same token, $d\Strain_{zz}=0.1$ is considered for the oedometric loading and $2\,d\Strain_{xz}=0.1$ for the simple shear loading.

\begin{table}[h]
    \caption{Mechanical parameters of the phases.}
    \centering
    \begin{tabular}{|l|l|}
    \hline
    Young modulus of the undissolvable grain (Pa) & 1 \\
    Young modulus of the dissolvable cement (Pa) & 0.1-0.25-0.5-1\\
    Young modulus of the pore (Pa) & 0\\
    Poisson ratio of the grain and the cement (-) & 0.3\\
    Poisson ratio of the pore (-) & 0\\
    \hline
    \end{tabular}
    \label{Parameters PF}
\end{table}

\begin{figure}[h]
    \centering
    \includegraphics[width=0.9\linewidth]{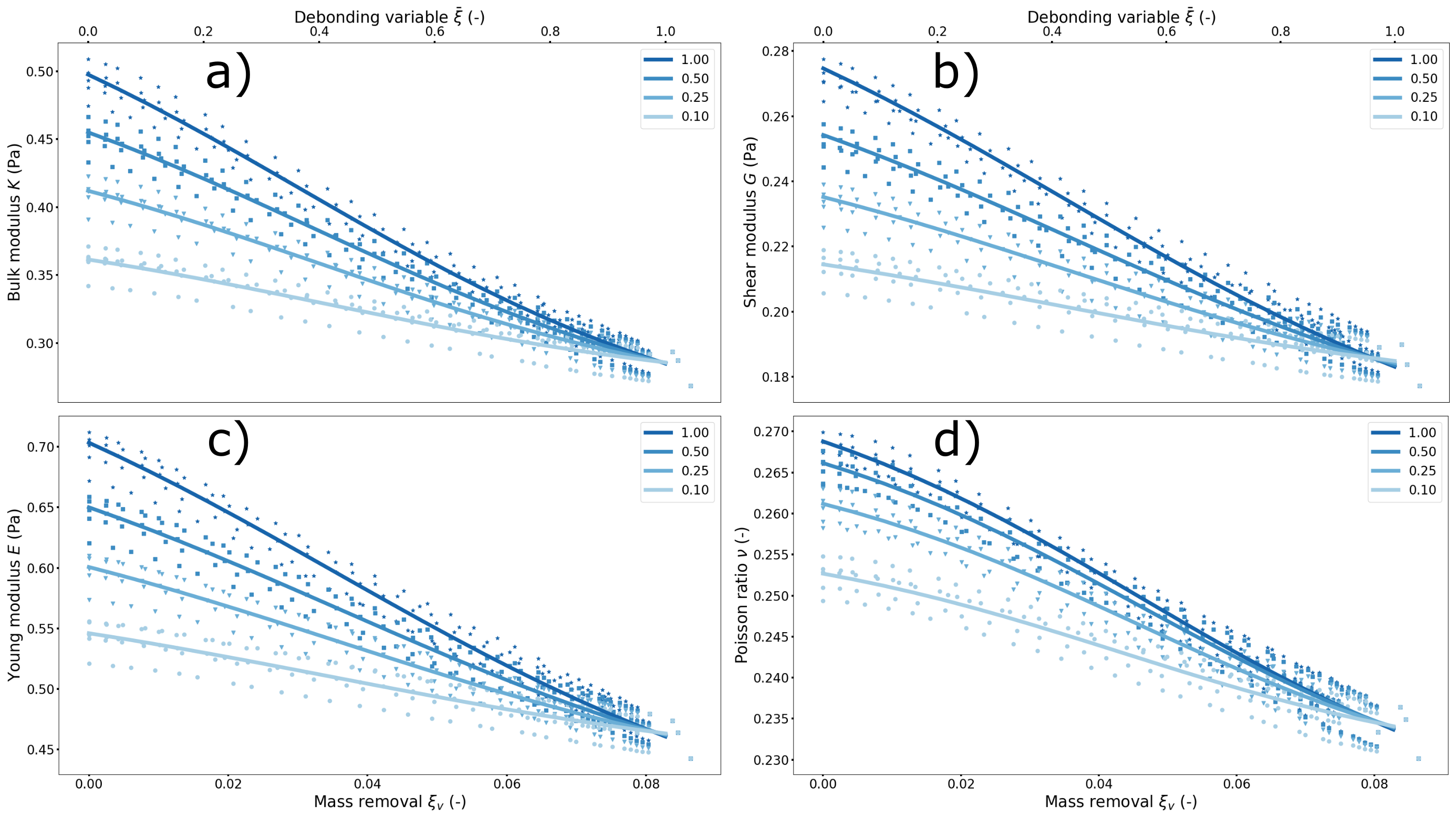}
    \caption{Evolution of a) the bulk modulus, b) the shear modulus, c) the Young modulus, and d) the Poisson ratio with the mass removal $\xi_v$ for distinct Young modulus for the cement material.}
    \label{Results Figure Pi 2}
\end{figure}

The isotropic loading test described in Subsection \ref{Isotropic Loading 2} is employed to determine the bulk modulus by applying Equation \ref{Bulk Interpolation} to the sample behavior.
Likewise, the oedometric loading test described in Subsection \ref{Oedometric Loading} is employed to estimate the Poisson ratio and the Young modulus by applying Equations \ref{Poisson Oedometric Interpolation} and \ref{Young 2 Interpolation} to the sample behavior.
Similarly, the simple shear loading test described in Subsection \ref{Shearing Loading} is employed to determine the shear modulus by applying Equation \ref{Shear Interpolation 2} to the sample behavior.
The evolution of these parameters is presented in Figure \ref{Results Figure Pi 2}.

As expected, the parameters decrease with the mass removal, and this reduction appears to follow a sigmoid-like decay. 
It is worth noting that the Poisson ratio decreases with the mass removal, in contrast to the tendency depicted in Figure \ref{Results Figure Pi}.
From the data available, it is difficult to determine if this fluctuation in the behavior is induced by the difference in the approach employed or by the difference in the cementation patterns between the samples.
Complementary investigations should be conducted to clarify this point.
We introduce the expression $\Pi(\xi_v)= \Pi_2 +\frac{\Pi_1-\Pi_2}{1+\text{exp}^{k_\Pi\left(\xi_v-\xi_{0,\,\Pi}\right)}}$, which fits accurately the simulation results, see Figure \ref{Results Figure Pi 2}, and the best-fit values of the coefficients $\Pi_1$, $\Pi_2$, $k_\Pi$, and $\xi_{0,\,\Pi}$ were determined by least square error fitting and are specified in Table \ref{Results Table Pi 2}. 
The validation domain of this fitted relation is restricted to $\xi_v\in\left[0,\;\xi_{v,\,max}\right]$.
If $\xi_v\geq \xi_{v,\,max}$, the sample property is considered equal to $\Pi(\xi_{v,\,max})$.
This parameter $\xi_{v,\,max}\; (=0.0828 \text{ herein})$ corresponds to the mean maximum mass removal possible from the microstructures employed to obtain the data points.
As a reminder, this data set is built on five separate microstructures, which evolve with the debonding phenomenon.

\begin{table}[h]
    \caption{Parameters estimated for the relation $\Pi(\xi_v)= \Pi_2 +\frac{\Pi_1-\Pi_2}{1+\text{exp}^{k_\Pi\left(\xi_v-\xi_{0,\,\Pi}\right)}}$ considering different cement stiffnesses.}
    \centering
    \begin{tabular}{|l||c|c|c|c||c|}
        \hline
        \multicolumn{6}{|c|}{Bulk modulus $K(\xi_v)$}\\
        \hline
        Young modulus of the cement (Pa) & $K_1$ (Pa) & $K_2$ (Pa) & $k_K$ (-) & $\xi_{0,\,K}$ (-) & $R^2$ \\
        \hline
        0.10 & 0.262 & 0.395 & -31.8 & 0.0346 & 0.895 \\
        0.25 & 0.247 & 0.470 & -31.5 & 0.0334 & 0.959 \\
        0.50 & 0.234 & 0.536 & -31.2 & 0.0321 & 0.976 \\
        1.00 & 0.220 & 0.606 & -30.8 & 0.0306 & 0.984 \\
        \hline
        \multicolumn{6}{|c|}{Shear modulus $G(\xi_v)$}\\
        \hline
        Young modulus of the cement (Pa) & $G_1$ (Pa) & $G_2$ (Pa) & $k_G$ (-) & $\xi_{0,\,G}$ (-) & $R^2$ \\
        \hline
        0.10 & 0.173 & 0.230 & -28.5 & 0.0348 & 0.825 \\
        0.25 & 0.164 & 0.261 & -28.3 & 0.0356 & 0.934 \\
        0.50 & 0.155 & 0.291 & -27.9 & 0.0356 & 0.964 \\
        1.00 & 0.145 & 0.324 & -27.6 & 0.0349 & 0.978 \\
        \hline
        \multicolumn{6}{|c|}{Young modulus $E(\xi_v)$}\\
        \hline
        Young modulus of the cement (Pa) & $E_1$ (Pa) & $E_2$ (Pa) & $k_E$ (-) & $\xi_{0,\,E}$ (-) & $R^2$ \\
        \hline
        0.10 & 0.428 & 0.592 & -27.1 & 0.0351 & 0.838 \\
        0.25 & 0.404 & 0.674 & -27.5 & 0.0358 & 0.937 \\
        0.50 & 0.382 & 0.750 & -27.5 & 0.0359 & 0.964 \\
        1.00 & 0.359 & 0.834 & -27.4 & 0.0354 & 0.976 \\
        \hline
        \multicolumn{6}{|c|}{Poisson ratio $\nu(\xi_v)$} \\
        \hline
        Young modulus of the cement (Pa) & $\nu_1$ (Pa) & $\nu_2$ (Pa) & $k_\nu$ (-) & $\xi_{0,\,\nu}$ (-) & $R^2$ \\
        \hline
        0.10 & 0.228 & 0.258 & -36.3 & 0.0432 & 0.909 \\
        0.25 & 0.225 & 0.268 & -35.6 & 0.0453 & 0.957 \\
        0.50 & 0.222 & 0.275 & -34.6 & 0.0462 & 0.969 \\
        1.00 & 0.221 & 0.279 & -33.7 & 0.0461 & 0.973 \\
        \hline
    \end{tabular}
    \label{Results Table Pi 2}
\end{table}

About the fitting relation, the general sigmoid function captures the reduction of the slope at the extrema of the domain ($\xi_v=0$ and $\xi_v=\xi_{v,\,max}$).
Indeed, the slope reduction at $\xi_v=0$ is explained by the fact that the cement dissolution does not perturb the sample initially. 
Even if the experimental procedures maximize the cement deposition at the contacts, where it participates in the force transmission \citep{Tengattini2023, Tengattini2023b}, some cement located around the grains does not participate in this force transmission.
This initial dissolution of the inactive cement minimizes the effect of the dissolution on the sample properties.
In the same vein, the slope reduction at $\xi_v=\xi_{v,\,max}$ is due to the fact that the contribution of the bonds is minimal close to the total dissolution of the cement in the sample.
This particular aspect was also observed in Section \ref{Results Section DEM} with the exponential curves.
As explained in Section \ref{DEM Section}, the reduction pattern depends strongly on the bond size distribution of the sample and on the localization of the cement material.

From Table \ref{Results Table Pi 2}, it appears that the coefficient $\Pi_1$ is quite independent of the intrinsic Young modulus for the cement. 
Indeed, it is related to the final value of the sigmoid function, which is common to all the simulations, being the unbonded state.
In the same way, the coefficient $\Pi_2$ increases with the Young modulus of the cement, as this parameter describes the amplitude (from the difference with the value $\Pi_1$) of the sigmoid function.
The slope of this function is governed by the coefficient $k_\Pi$, which remains mostly constant with the cement Young modulus.
Finally, the curve is translated following the x-axis by the coordinate of the inflection point $\xi_{0,\,\Pi}$, which varies slightly with the Young modulus of the cement phase.

The $R^2$ coefficients from the Table \ref{Results Table Pi 2} show that the fitted relations are good approximations within the range of our numerical examples.
However, it is important to point out that a single microstructure gives much lower values than the others, reducing the $R^2$ coefficients.
Even if the dimension of the domain considered ensures that a Representative Elementary Volume is studied, see Appendix \ref{REV PF}, this specific microstructure may be a particular sub-domain involving a influent heterogeneity.
These data have been kept to showcase the potential limitation of the method depicted herein.
The influence of such a singularity can be reduced by multiplying the number of samples considered \citep{Loyola2021,Zwartz2024}.

\vskip\baselineskip

The quality of the relations $\Pi(\xi_v)$ is now evaluated. First, one should ensure that the end member (corresponding to the unbonded state) of the approximation functions gives similar properties. Moreover, the Young modulus of the cement should have no influence on the sample properties at large mass removal, as the cement material is entirely dissolved.
Table \ref{Stats Interpolated Unbonded State 2} depicts the minimal, the maximal, and the mean values of $\Pi(\xi_v\rightarrow\infty)=\Pi(\xi_v=\xi_{v,\,max})$, considering the different cement stiffnesses.
These elastic parameters for the approximated unbonded states appear consistent, with a maximum relative error of 0.5\% compared to the mean value.

\begin{table}[h]
    \caption{Variation of the elastic parameters for the approximated unbonded state $\Pi_{\infty}=\Pi(\xi_v=\xi_{v,\,max})$.}
    \centering
    \begin{tabular}{|l||r||c|c|c|}
        \hline 
        $\Pi$ & 0.10-0.25-0.50-1.00 values & min & mean & max \\
        \hline 
        K (Pa) & 0.2857-0.2855-0.2852-0.2849 & 0.2849 & 0.2853 & 0.2858 \\
        G (Pa) & 0.1848-0.1842-0.1836-0.1830 & 0.1830 & 0.1839 & 0.1848 \\
        E (Pa) & 0.4631-0.4624-0.4617-0.4606 & 0.4606 & 0.4620 & 0.4631 \\
        $\nu$ (-) & 0.2341-0.2340-0.2339-0.2337 & 0.2337 & 0.2339 & 0.2341 \\
        \hline
    \end{tabular}
    \label{Stats Interpolated Unbonded State 2}
\end{table}

\begin{figure}[h]
    \centering
    \includegraphics[width=0.9\linewidth]{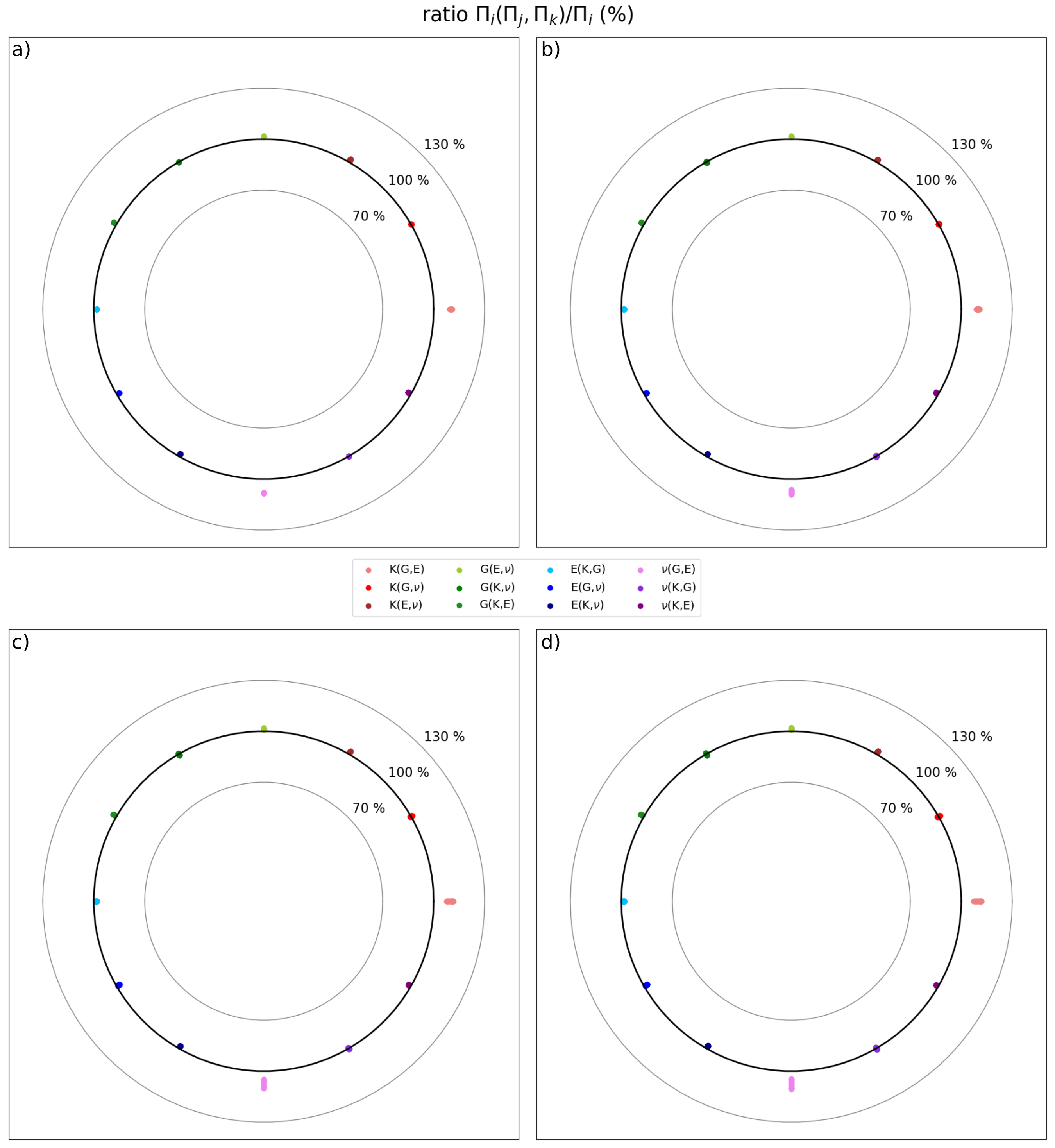}
    \caption{Cross verification of the estimated parameters for the Young modulus of the cement equals a) 0.10, b) 0.25, c) 0.50, and d) 1.00 Pa.}
    \label{Results Figure Cross Verification 2}
\end{figure}

As described in Equations \ref{Equation K}-\ref{Equation v}, the distinct elastic parameters are correlated.
Similar to Figure \ref{Results Figure Cross Verification}, the goal is to verify the parameter estimations with the relation $\Pi_i$ vs. $\Pi_i(\Pi_j,\,\Pi_k)$, where $\Pi$ is the bulk modulus, the shear modulus, the Young modulus, or the Poisson ratio.    
In particular, the Figure \ref{Results Figure Cross Verification 2} depicts the ratio between the determined parameter and the estimated parameter $\Pi_i$ vs. $\Pi_i(\Pi_j,\,\Pi_k)$ for different mass removal $\xi_v$. The different points must be located on the isoline ratio = 100\%.

The relations between the elastic parameters are verified as the points are located around the 100\% isoline.
However, some fluctuations can be observed for $\nu(Y,\,G)$ and $K(Y,\,G)$.
Like for the DEM study, the uncertainties of the moduli propagate in the relations depicted in Equations \ref{Equation K}-\ref{Equation v}.
Moreover, the shear modulus $G$, the Young modulus $E$, and the Poisson ratio $\nu$ have been obtained through simulations (oedometric and shear) by considering only one direction for the loading application. 
However, the microstructure investigated is slightly anisotropic, induced by the extraction of the subset.
The quality of the estimation of these parameters can be increased by multiplying the simulations.
For instance, the oedometric loading can be applied in the x-, y-, and z-axis.
Similarly, the shear loading can be applied in x- and y-axis considering the z-normal face, in the y- and z-axis considering the x-normal face, and in the x- and z-axis considering the z-normal face.
Doing so, the shear modulus, the Young modulus, and the Poisson ratio of the sample may be obtained with an average value of these estimations.
Moreover, this average value would be more consistent with the isotropy assumed in Equations \ref{Equation K}-\ref{Equation v}.
It is worth noting that the bulk modulus is more certain, as the loading is isotropic, and the average over the distinct directions is conducted by definition.
However, the quality of the fitted relations remains accurate compared to the ones depicted in Section \ref{Results Section DEM}.


\section{Discussion}
\label{Discussion Section}

\subsection{On the use of the chemical degradation laws}
\label{Use Discussion Section}

The main advantage of the framework detailed in this contribution is to incorporate micro-mechanisms into the estimation of the sample behavior.
Such an approach has already been applied in the literature.
For instance, \emph{Gajo et al.} established a weathering function, assuming a Representative Elementary Volume composed of eight grains that are organized in a cubic mesh \citep{Gajo2015, Gajo2019}, limiting the application to real samples.
In the same vein, \emph{Buscarnera and Das} related the property alteration to an evolution of the bond size distribution, involving parameters that can be difficult to calibrate. 
In parallel with facing these difficulties, the proposed approaches introduce additional micro-mechanisms, such as an explicit description of the organization between the phases.  
This aspect is fundamental, as it has been emphasized that the local porosity, which lies at the microscale, dictates the mechanical deformation field at the microscale \citep{DoreOssipyan2025}.
In the same tone, distinct microstructures can be generated with similar geometrical characteristics, inducing potential fluctuations in the behavior of the sample \citep{Lesueur2020b,OlarteGarzon2027}.
The digital twin framework employed herein reduces these uncertainties concerning the organization of the phases.
The consideration of these micro-mechanisms implies the establishment of more complex alterations of the effective properties than the linear dependency often employed in the literature.
Furthermore, this method can be easily extended to analyze the behavior at the grain scale or to explore the effect of additional solicitations.

\vskip\baselineskip

Nevertheless, it is pivotal to limit the validity of these relations to the hypothesis made in this contribution. 

First, the sample is assumed herein to remain in the elastic domain, restricting the origin of damage to the chemical reduction of the cement and neglecting potential additional damage due to cement breakage.
For instance, this mechanical rupture is incorporated in the softening laws through an additional damage variable (dedicated to the mechanical alteration) \citep{Schuler2020} or with a modification of the damage variable employed (representing the chemical and mechanical alterations simultaneously) \citep{Wu2026}.
Such a relation could be informed by the framework depicted in this contribution by activating the bond breakage. 
Thus, one could conduct an investigation to determine the dependence of the tangent elastic properties on the load increment amplitude, which can be chemical, mechanical, or a mix. 

Second, these laws have been formulated from a three-phase sample (granular skeleton, localized cementation, pore) during the debonding phenomenon. 
Consequently, they can not be applied to more complex microstructures or to materials subjected to different weathering phenomena (such as grain and cement dissolution). 

Third, the softening laws were evaluated on a limited number of numerical examples.
Grain and bond size distribution, the grains and cement properties, or the organization of the phases may differ substantially from specific rocks. 
New weakening relationships may be determined for such materials by using our numerical workflow.


\subsection{The discrete approach, the continuous approach, or a combination of the two}
\label{DEM vs PF}

As mentioned previously, the samples investigated with the discrete approach (Section \ref{Results Section DEM}) and the continuous approach (Section \ref{Results Section PF}) differ.
The difference between the cementation mechanisms (biocementation versus capillary bridges) results in different local cement deposition, see Figure \ref{Figure biocementation capillary bridge}.
Thus, the direct comparison between the numerical methods employed is difficult to make.
In particular, the origin of the variation in the evolution of the Poisson ratio (increasing for biocementation and decreasing for capillary bridges) remains uncertain.
More than the numerical method employed, the alteration law for the evolution of the Poisson ratio depends strongly on the initial cementation mechanism. 
Concerning the other moduli (bulk, shear, and Young), the decreasing tendencies remain similar for the two approaches.

The two methods predict a slope reduction at the end of the debonding phenomenon, and only the continuous approach estimates a slope reduction at the beginning of the weathering.
Concerning the first point (slope reduction at the end), this behavior is representative of the reduction in importance of the bonds in the composite sample, as predicted by a simple system composed of two springs in parallel (one for the cement and one for the granular skeleton).
After a certain alteration state, the equivalent stiffness of the system is dictated by the granular skeleton spring, and the participation of the cement spring is negligible.
Concerning the second point (slope reduction at the beginning), this phenomenon is due to the dissolution of the inactive cement material (that does not participate in the stress transmission). 
This quantity of inactive cement material depends strongly on the cementation pattern in the microstructure.
Thus, this ability of the numerical method could be required to correctly estimate the alteration of the sample properties.
However, this phenomenon could be negligible if the cement material is mainly located in active spots.
Furthermore, the mass removal $\xi_v$ is much easier to estimate during experimental tests than the cement surface reduction $\xi_s$, facilitating the establishment of a synergy between the numerical methods depicted herein and experimental campaigns.

\vskip\baselineskip

As depicted in Section \ref{DEM Section}, the DEM is a promising approach to investigate the behavior of a cemented granular material while debonding occurs. 
In particular, this approach describes in an explicit manner the organization and the interactions of the grains.
However, the main restriction of this formulation is the fact that the cement phase, located between the grains, is not explicitly described. 
Thus, a segmentation method is required, see an example in \citep{Tengattini2023}, to determine from XRCT-scans the bond area associated with each cemented contact.
It is essential to remember that this 2D geometrical dimension is estimated from a 3D configuration, requesting modeling assumptions.
In the same note, a cement point can connect more than two grains, increasing the difficulty of identifying the geometrical characteristics of these distinct contacts.
Moreover, the reduction of the cement matter employed herein is a homogeneous dissolution.
This assumption does not verify the fact that the dissolution kinetics depend on the specific surface, and so on the shape and size of the material \citep{Rattez2021}, on the stress state \citep{SacMorane:PFDEM, SacMorane:PFDEMb, Guevel2020}, and on the propagation of the reactive species \citep{Li2023, Xie2026}.

Facing these difficulties, our continuous approach allows a physics-based prediction of the dissolution pattern and the explicit consideration of the cement in the material, see Section \ref{PF+FFT Section}.
However, the use of such a method does not capture some fundamental phenomena at the microscale, such as granular reorganization. 
For instance, it has been emphasized by \citep{SacMorane:StressState} that the collapse of chain force dictates the evolution of the sample stress state during the debonding phenomenon. 
The ability of the DEM to capture such a mechanism remains essential.

Recently, a novel PFDEM method has been formulated, coupling the PF description and the DEM \citep{SacMorane:PFDEM, SacMorane:PFDEMb}.
In particular, it appears that this framework captures (1) irregular grain shapes, (2)  granular reorganization, (3) heterogeneous dissolution/precipitation, (4) force transmitted at the contacts, (5) diffusion of the solute in the pore space, and (6) rate-limiting processes.
This advanced model appears to carry the advantages of the two methods explored in this contribution.
However, the remaining limitation is the computational cost. 
If the duration of an individual simulation depicted in Section \ref{Results Section} is a couple of hours on a regular computer (Intel Xeon E5-2630 v3 x 32 and 48 Gb of memory), the simulation time for similar configurations is weeks to months with a PFDEM approach, in particular for 3D simulations. 
The PF resolution appears to be the bottleneck of the computational cost \citep{SacMorane:PFDEM}.
This limitation dictates the number of grains in the subset investigated, threatening its representativeness. 
However, numerous methods may be employed to diminish the resources required, allowing the use of the PF description for 3D problems, considering multi-physics \citep{Hu2001,Sessim2022, Battas2025}. 
Moreover, an under-representative volume can be considered for investigation by multiplying the repetition of the framework and averaging the obtained values \citep{Loyola2021, Zwartz2024}.
Future work should apply a 3D PFDEM method to more accurately capture the driving mechanisms during the debonding, adding the explicit description of the cement to the observations depicted in Section \ref{Results Section DEM} and the granular reorganization to the observations depicted in Section \ref{Results Section PF}.


\section{Conclusion}

This contribution investigates the evolution of elastic parameters of cemented granular materials in the context of weathering.
In particular, this paper focuses on the effect of the debonding phenomenon, defined as the reduction of the cement while the material constituting the granular skeleton remains intact.
Two numerical workflows were developed: first, the discontinuous approach uses a Discrete Element Model for geometry construction and effective elastic property determination, and represents weathering through a scalar reduction of bond stiffness. Second, the continuous approach employs a Phase-Field formulation for 3D bond dissolution, combined with the Fast Fourier Transform to compute effective elastic properties.
Both approaches were applied in numerical examples on cemented granular materials from the literature.
In these examples, our approach is able to determine effective softening laws, that describe the change of effective properties with evolving debonding.
Fundamental properties for these descriptions, such as the common unbonded state and the relations between the elastic parameters, have been verified.
Compared to the existing models, the methods employed have the advantages of considering the real organization of the phases and of being easier to calibrate.

In particular, the use of such approaches reveals mechanisms involved at the microstructure level during the debonding phenomenon. 
For instance, the Discrete Element Method has emphasized the slowdown of the decay of the softening law with the increase of the bond surface reduction. 
Indeed, the presence of the bond becomes negligible while its volume fraction reduces, explaining the lower effect on the sample properties.
Likewise, the combination of the Phase-Field formulation with the Fast Fourier Transform has highlighted a reduction of the slope at the start of the weathering phenomenon.
Indeed, this observation is explained by the presence of inactive cement in the microstructure. 
This notion of inactive cement is relative to the fact that this element does not participate in the transmission of stresses.

Even if the behaviors estimated in this work are based on physical considerations, experimental campaigns should be conducted in the future to challenge them.
Thus, the difficulties would be to investigate a representative cemented granular material that undergoes the debonding phenomenon in a reasonable time duration.
In the same tone, the framework depicted herein could be easily extended to consider additional multiphysics solicitations and to integrate complementary micro-mechanisms, such as bond breakage.
Similarly, coupling the advantages of the two methods employed, the PFDEM method is a promising approach for the tackled problem.
However, the related computational cost should be reduced in order to study representative microstructures.
Finally, further investigations on the dependency of these softening laws should be conducted. In particular, the size distribution of the grain and the cement, the shape of elements, or the intrinsic parameters are strong candidates to significantly influence these softening laws.


\section*{Acknowledgements}

This research has been partially funded by the Fonds Spécial de Recherche (FSR), Wallonia-Bruxelles Federation, Belgium. The work has also received funding from the National Science Foundation (NSF), USA, project CMMI-2042325. The present work is developed within the framework of the REFROZEN ANR-DFG project with the Aachen University of Technology (RWTH) in Germany. This work has benefited from funding by the French National Research Agency (Grant ANR‐22‐CPJ1‐0027‐01).

We would like to thank A. Tengattini for sharing the XRCT-scans and F. Wu for motivating this investigation.


\appendix

\section{The lognormal distribution}
\label{Lognormal Distribution}

The probability of a bond to have a surface $A_b$ follows a lognormal distribution formulated in Equation \ref{Lognormal Distribution Equation}. This distribution is defined by the expected value $m_{log}$ and the variance $s_{log}$.

\begin{equation}
    p(A_b) = \frac{1}{A_b\,s_{log}\sqrt{2\pi}}e^{-\frac{\left(ln\left(A_b\right)-m_{log}\right)^2}{2\,s_{log}^2}}
    \label{Lognormal Distribution Equation}
\end{equation}
    
It appears that this distribution reproduces accurately experimental observations \citep{Sarkis2022}. Indeed, smaller bond surfaces and larger ones can be considered; both are crucial to the mechanical behavior of the sample. 


\section{Variation in the configuration of reference}
\label{ev0}

\begin{figure}[h]
    \centering
    \includegraphics[width=0.7\linewidth]{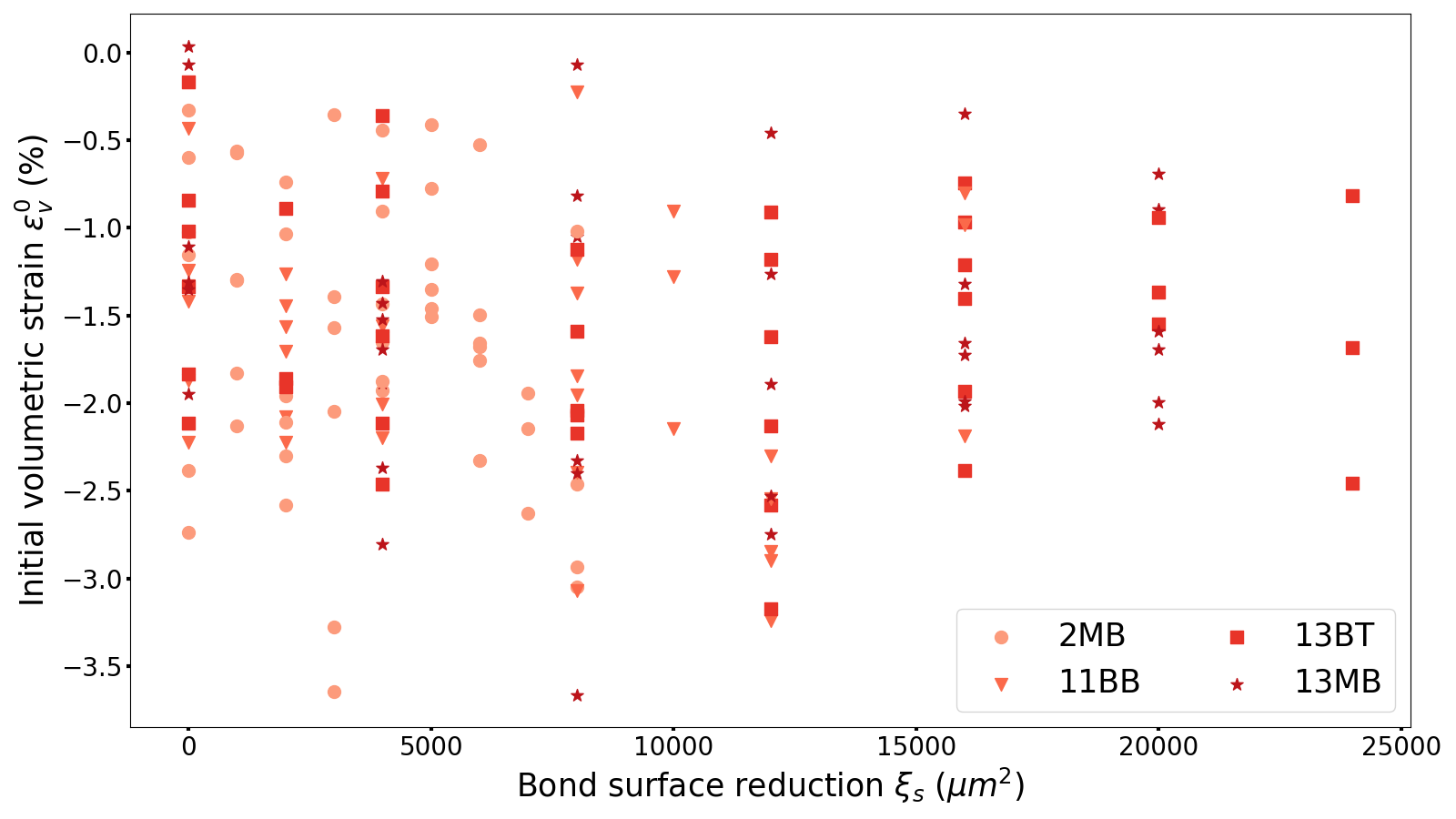}
    \caption{Fluctuation of the volumetric strains once the sample is generated, before reinitializing the reference.}
    \label{Results Figure ev0}
\end{figure}

As explained in Section \ref{Results Section DEM}, the configuration of reference may vary between the samples investigated due to the sample preparation algorithm depicted in Figure \ref{Initial condition algorithm}.
This variation is illustrated with the volumetric strain $\epsilon_v^0$ obtained at the end of the initialization (step f of Figure \ref{Initial condition algorithm}), see Figure \ref{Results Figure ev0}.
It appears that this variance remains negligible among the samples investigated.

\section{Determine the Representative Elementary Volume of a cemented granular medium}
\label{REV PF}

\begin{figure}[h]
    \centering
    \includegraphics[width=0.7\linewidth]{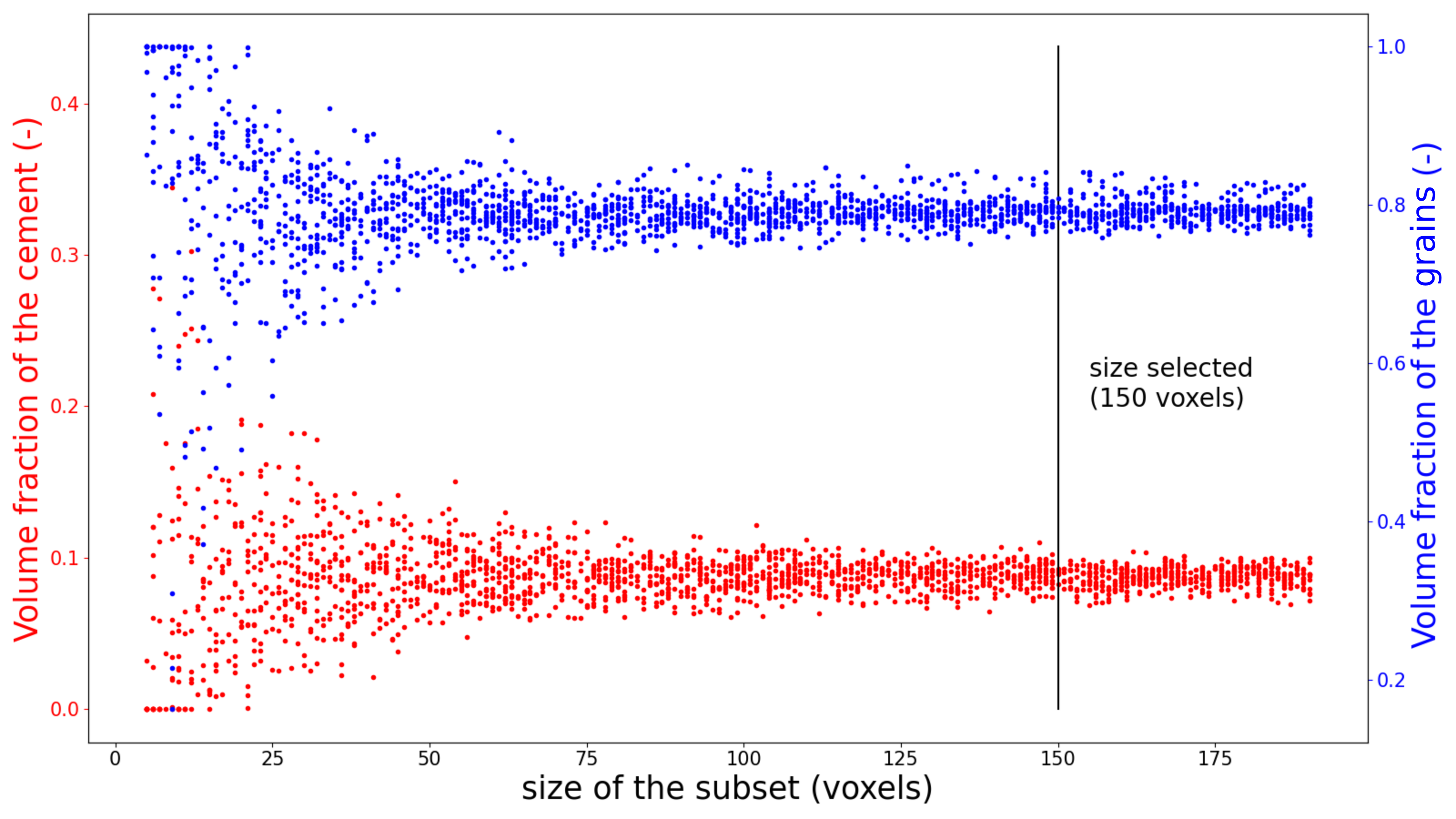}
    \caption{Cement and grains volume fractions with different subset dimensions.}
    \label{Results Figure REV}
\end{figure}

The determination of the Representative Elementary Volume is based on a morphometer: the volume fraction of the distinct phases (grain, cement, and pore).
Figure \ref{Results Figure REV} depicts the value of these morphometers for distinct subsets extracted from the XRCT-scan of the sample. 
In particular, it appears that the variation of the parameters decreases with the dimension of the subset. 
The Representative Elementary Volume can then be determined by considering a size that minimizes the variation. 
Moreover, the repetition of the parameter estimation allows the determination of the representative value by averaging the obtained values \citep{Loyola2021, Zwartz2024}.
Figure \ref{Results Figure REV} justifies the selection of a 150x150x150 voxels subset in Section \ref{PF+FFT Section}.

\printcredits

\bibliography{bibliography}
\bibliographystyle{ieeetr}

\end{document}